\documentclass[aps,prb,showpacs,reprint, superscriptaddress]{revtex4-1}

\usepackage{graphicx}
\usepackage{color}
\usepackage{amsmath}
\usepackage{bbm}
\usepackage{amssymb}
\usepackage{braket}
\usepackage{dsfont}

\usepackage[utf8]{inputenc}	% kodowanie UTF-8
\usepackage[T1]{fontenc}

\input epsf

\begin{document}

\title{Honeycomb anti-dot artificial lattice\\ as a prototypical correlated Dirac fermions system}% Force line breaks with \\
\author{A. Biborski}
\email{andrzej.biborski@agh.edu.pl}
\affiliation{Academic Centre for Materials and Nanotechnology, AGH University of Krakow, Al. Mickiewicza 30, 30-059 Krakow,
Poland}

\author{M. Zegrodnik}
\email{michal.zegrodnik@agh.edu.pl}
\affiliation{Academic Centre for Materials and Nanotechnology, AGH University of Krakow, Al. Mickiewicza 30, 30-059 Krakow,
Poland}

\author{P. Wójcik}
\email{pawel.wojcik@fis.agh.edu.pl}
\affiliation{Faculty of Physics and Applied Computer Science, AGH University of Krakow, Al. Mickiewicza 30, 30-059 Krakow,
Poland}

\author{M. P. Nowak}
\email{mpnowak@agh.edu.pl}
\affiliation{Academic Centre for Materials and Nanotechnology, AGH University of Krakow, Al. Mickiewicza 30, 30-059 Krakow,
Poland}

\begin{abstract}
We study theoretically the electronic properties of the artificial quantum dot honeycomb lattice defined in a two-dimensional electron gas, focusing on the possibility of achieving a regime in which electronic correlations play a dominant role. At first we establish a non-interacting model compatible with recently studied experimentally devices. According to the values of the obtained electron-electron interaction integrals, we postulate that the inclusion of inherent electron-gas self-screening is indispensable to reconstruct the experimental observations. Applying the Thomas-Fermi type of screening, we show that the radius of the anti-dot is crucial to achieve a correlated state in which phenomena like antiferromagentic ordering and interaction-induced insulating state appear. We estimate the conditions for which the electronically correlated state in an artificial honeycomb lattice can be realized.
\end{abstract}

\maketitle

%\tableofcontents

\section{\label{sec:introduction} Introduction}
There is a long-standing debate about the role of electron-electron interactions in two-dimensional Dirac fermion systems~\cite{Kotov,hokin,DasSarma2007}. In particular, for graphene, massless fermionic quasiparticles are believed to be strongly renormalized by the presence of electronic interactions~\cite{Kotov}. However, this renormalization influences the Fermi velocity and thus keeps the semi-metallic character of the system instead of inducing the Mott insulator phase. The unscreened long-range interactions in two-dimensional systems may also play a competitive role with respect to Mott localization~\cite{hokin,WuWei,Assad,Sorella2012}. However, short-range Hubbard interactions $U$ are believed to be responsible for the formation of a gap in strongly correlated systems when the kinetic energy scale given by the hopping amplitude $|t|$ is noticeably smaller than $U$. In the absence of long-range interactions, the gap opening in the Dirac fermion systems is supposed to arise for $U/|t| \gtrsim 3.8\div4.2$~\cite{Meng2010,Sorella2012,Raczkowski}. This quantum phase transition from the semi-metal to the antiferromagnetic insulator at half-filling has been pointed out in a number of theoretical studies\cite{hokin,WuWei,Otsuka,Liebsch,Assad,Meng2010,Sorella2012,Arya,Raczkowski}.

The electronic properties of honeycomb systems can be conveniently studied using the so-called \emph{quantum simulators} such as artificial lattices of quantum dots (ALDs) defined in a two-dimensional electron gas (2DEG). This approach provides the possibility of tuning the properties of the system by controlling the size of the dots and the spacing between them, as well as the depth of their confinement potential~\cite{Saleem}. The emergence of Dirac systems in ALD has been experimentally demonstrated by means of photoluminescence measurements in a nanodevice constituted in the so-called anti-dot (AD) architecture~\cite{DuLiu}. The AD approach provides the opportunity to form quasi-2D bulk systems, contrary to quantum dot assemblies fabricated by electrostatic gating, where the number of trapping centers typically does not exceed a dozen. Nevertheless, the photoluminescence spectra presented by Du et al.~\cite{DuLiu} indicate the presence of the spectral dublet related to excitations associated with van Hove singularities, which maps accurately on the model of non-interacting electrons. Thus, at least for the AD radius and lattice spacing that was examined, this device cannot be regarded as a strongly correlated system. Therefore, the natural question arises~\cite{Saleem}, if tuning of these two parameters may result in forcing ALD to be in the correlated state, characterized, e.g., by formation of the Mott gap, emergence of antiferromagnetic (AF) order, etc. The artificial honeycomb lattices formed in 2DEG have been theoretically studied in view of their electronic properties by a priori assuming an idealized confinement similar to graphene~\cite{Kylanpaa,TommyLi, Rasanen} or ADs~\cite{Saleem,Krix,Yun-Mei}. These valuable pioneering papers provide evidence of the influence of electron-electron interactions on the ALD band structure in the framework of Density Functional Theory ~\cite{Rasanen}, possible mechanism of pairing~\cite{TommyLi}, or the stability of the Dirac cone with respect to the shape of the confining potential~\cite{Kylanpaa} or disorder~\cite{Tkachenko_2015}. 

In this work, we investigate conditions for realization of a correlated state in an artificial Dirac lattice in realistic structures. For concreteness, we take the device as recently realized by Du et al.~\cite{DuLiu} as a starting point and investigate the electronic properties of 2DEG ALD, establishing the conditions under which those systems transit into the correlated phase. We present the AD-ALD model elaborated within multiscale simulations. Namely, first, we construct the mean-field model exploiting the Schr\"odinger-Poisson scheme. Subsequently, we utilize the resulting single-particle picture for construction of the Wannier basis by means of the projection method and subsequently calculate the electron-electron interaction amplitudes. We discuss the role of electronic screening, which we include by applying the Thomas-Fermi model with the screening length characteristic for GaAs-based 2DEG. Inclusion of screening is indispensable for the reconstruction of experimental finding, as bare (non-screened) amplitudes would imply a strongly correlated system for the lattice spacing and AD radius for which a semimetallic character has been reported~\cite{DuLiu}. Eventually, we elaborate the interacting Hamiltonian in the second quantization formulation and solve it for the half-filled case by means of the Variational Monte Carlo (VMC) method. These simulations provide evidence of the transition from the semimetallic state to the AF phase and indicators of the emergence of a Mott gap with an increase in AD diameter.

\section{\label{SCF}Model and Mean Field approach}

\subsection{Device and AD periodic potential}

\begin{figure}
\includegraphics[width=0.40\textwidth]{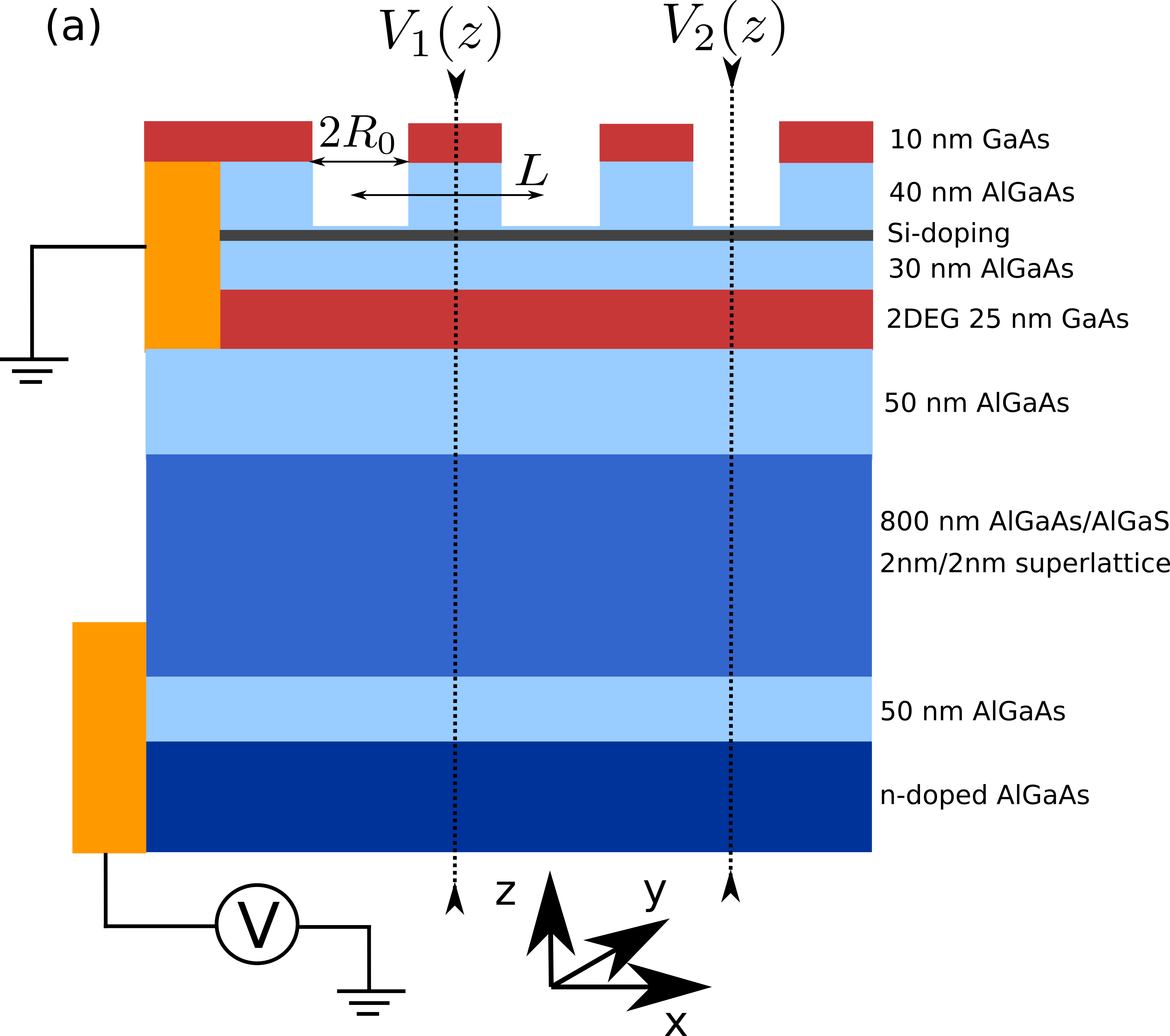}
\includegraphics[width=0.40\textwidth]{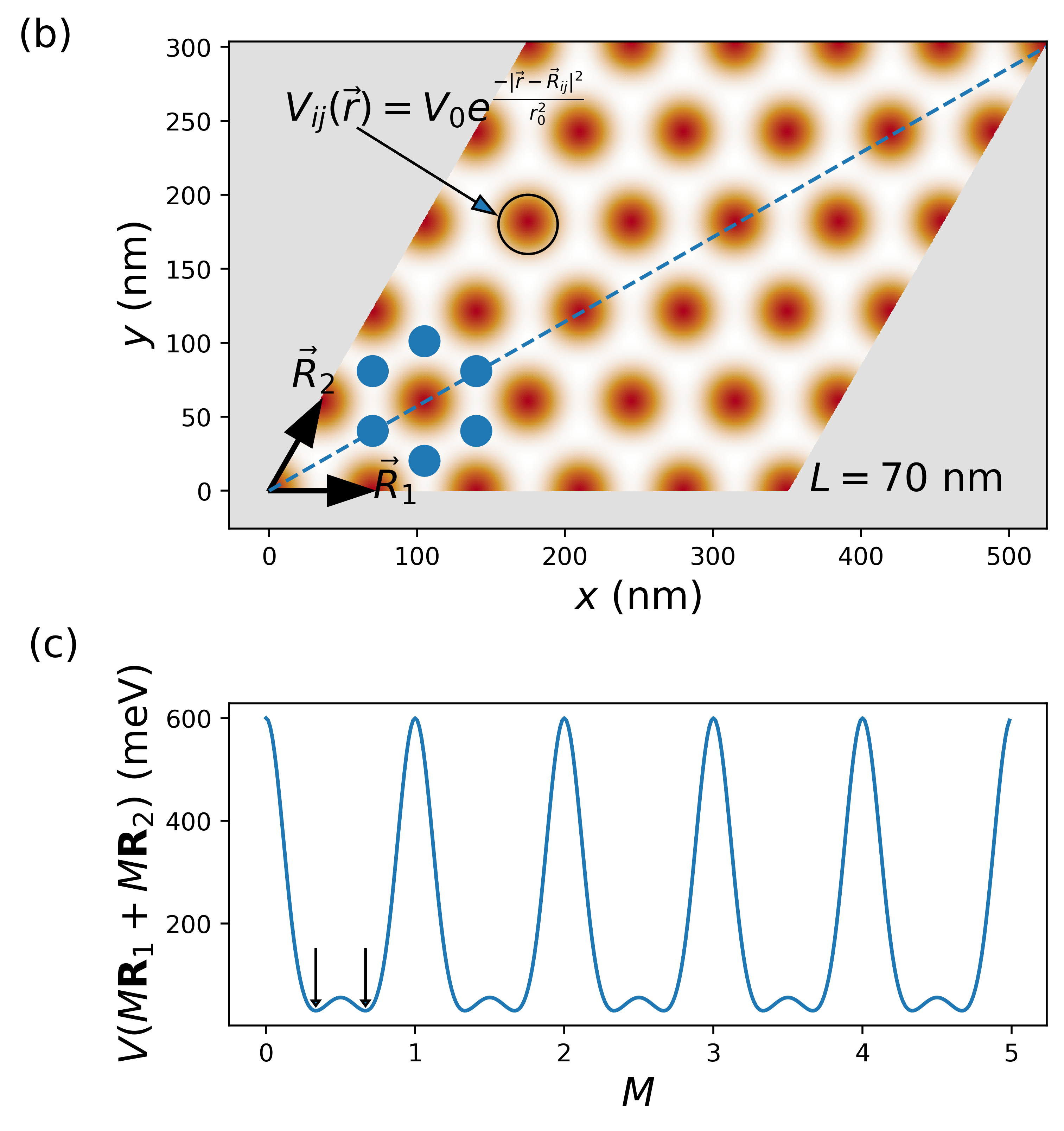}
\caption{(a) Cross section of the considered heterostructure. (b) Two dimensional plot of electrostatic potential induced by etching in 2DEG. Filled blue circles indicate examplary locations of potential minima.(c) The electrostatic potential profile in 2DEG along the direction marked as dashed line in (b). Arrows refer to the potential minima.}
\label{fig:device}
\end{figure}

We consider an artificial honeycomb lattice defined in a GaAs/AlGaAs heterostructure~\cite{DuLiu} (Fig. \ref{fig:device}(a)). ADs are assembled in a triangular lattice (Fig.\ref{fig:device}(b)) spanned by vectors $\mathbf{R}_1=(L,0)$ and $\mathbf{R}_2=(L/2,L\sqrt{3}/2)$ where $L=|\mathbf{R}_1|=70$ nm. The points related to $\mathbf{R}_{ij}=i\times \mathbf{R}_1+ j\times\mathbf{R}_2$ are associated with maxima of potential $V_{r_0}(\mathbf{r})$ resulting from the patterned etching of the top layers of GaAs/AlGaAs. Here, we model the periodic potential assuming a Gaussian contraction, which is given as
\begin{align}
    V_{r_0}(\mathbf{r})=\sum_{i,j}V_0\exp{\Bigg[-\Bigg(\frac{||\mathbf{R}_{ij}-\mathbf{r}||}{r_0}\Bigg)^2\Bigg]},
\label{eq:V}
\end{align}
where $\mathbf{r}=(x,y)$, $V_0$, and $r_0$ are related to the maximum height of the potential and the radius of the antidot, respectively. The Gaussian functional form is believed to properly describe the trapping potential that comprises quantum dots of diameter less than $\sim100$ nm in 2DEG formed in GaAs/AlGaAs heterostructures~\cite{Bednarek2,Ciurla}. As the estimated radius of the etched antidots in the experiment~\cite{DuLiu} is $r_0\approx20\pm5$ nm, we inspect the values of $r_0$ in a similar range. Note that since $r_0\approx L/3$, the local minima in the landscape of $V_{r_0}(\mathbf{r})$ present in Fig.~\ref{fig:device}(c) correspond to dots that form a graphene-like lattice.

The potential amplitude $V_0$  in Eq.~(\ref{eq:V}) is determined by means of the standard Schr\"odinger-Poisson approach used separately for two cross sections depicted in Fig.~\ref{fig:device}(a) by dashed lines. Assuming translation symmetry in the $x-y$ plane, $\Psi(\mathbf{r})=\phi_n(z)\exp (ik_xx+ik_yy)$, the Schr\"odinger equation can be reduced to the 1D form 
\begin{align}
\left ( -\frac{1}{2} \frac{d}{dx} \frac{1}{m^*_{\perp}(z)}\frac{d}{dx} + V_{r_0}(z) + \frac{\hbar ^2 k^2}{2m^*_{\parallel}} \right ) \phi _n(z)=E_n\phi _n(z),
\label{eq:schrodinger1D}
\end{align}
where $k^2=k_x^2+k_y^2$, $m^*_{\perp (\parallel)}$ is the effective mass in the direction perpendicular (parallel) to the layers and the potential $V_{r_0}(z)$ is the sum of two components: (i) $V_{band}(z)$ related to the discontinuity of the conduction band at the GaAs/AlGaAs interfaces and (ii) the electrostatic potential $V_e(z)$ that includes the electron-electron interaction and the electric field from the gate. The latter can be determined at the mean-field level from the Poisson equation given by 
\begin{align}
\epsilon_0\frac{d}{dz}\epsilon_r(z) \frac{d}{dz} V_e(z)=-\rho(z),
\label{eq:poisson1D}
\end{align}
with the charge density $\rho(z)=\rho_d(z)+\rho_e(z)$, where $\rho_d(z)$ is the doping profile and $\rho_e(z)$ is the electron distribution.
The electron density is obtained by
\begin{align}
\rho_e(z)=2\sum _n |\phi_n(z)|^2 f_{2D}(E_n - \mu),
\end{align}
where the factor $2$ accounts for the spin degeneracy, $T$ is the temperature, $\mu$ is the chemical potential and $f_{2D}$ is the Fermi-Dirac distribution integrated over the $k_x$ and $k_y$ components of the wave vector,
\begin{align}
f_{2D}(E_n,\mu)=\frac{m^*_{\parallel}k_B T}{2\pi} \ln \left  ( 1+\exp \left ( -\frac{E_n-\mu}{k_BT} \right ) \right ). 
\end{align}
Equations (\ref{eq:schrodinger1D}) and (\ref{eq:poisson1D}) are solved numerically using the finite difference method with Dirichlet boundary conditions. 
The existence of 2DEG in the main quantum well results from the Si delta doping located $30$ nm above. The value of $n_d$ is chosen to correspond to the electron concentration in 2DEG at the level of $n_{el}\approx 4.5\times 10^{10} \text{ cm}^{-2}$, i.e., about two electrons per unit cell, as provided in the experiment~\cite{DuLiu}. Then, for a chosen $n_d$ the self-consistent energy profile $V_2(z)$ is determined for the cross section with the etched region (see Fig.~\ref{fig:device}(a)) and the potential amplitude $V_0$ is estimated based on the adiabatic approximation according to the formula
\begin{align}
\label{eq:v0}
V_0\approx E_2^1 -E_1^1,
\end{align}
where $E_2^1$ and $E_1^2$ are energies related to the lowest lying states for the etched and non-etched cases, respectively. The procedure is repeated until self-consistency is reached, which we consider to occur when the potential variation between two consecutive iterations is less than $10^{-7}$ eV. 
Calculations have been carried out for the material parameters corresponding to GaAs (AlGaAs): $m^*_{\perp}($GaAs$)=0.067$, $m^*_{\perp}($Al$_x$Ga$_{1-x}$As$)=0.067+0.083x$, $m^*_{\parallel}($GaAs$)=m^*_{\parallel}($Al$_x$Ga$_{1-x}$As$)=0.067$, $\epsilon_r ($GaAs$)=13.18$, $\epsilon_r($Al$_x$Ga$_{1-x}$As$)=13.18+3.12x$ and the conduction band minima $E_c=E_g/2$, where the energy gap $E_g($GaAs$)=1.43$~eV, $E_g($Al$_x$Ga$_{1-x}$As$)=1.43+1.247x$. We set $x=0.3$ and $T=4$~K. 
\begin{figure}
\includegraphics[width=0.50\textwidth]{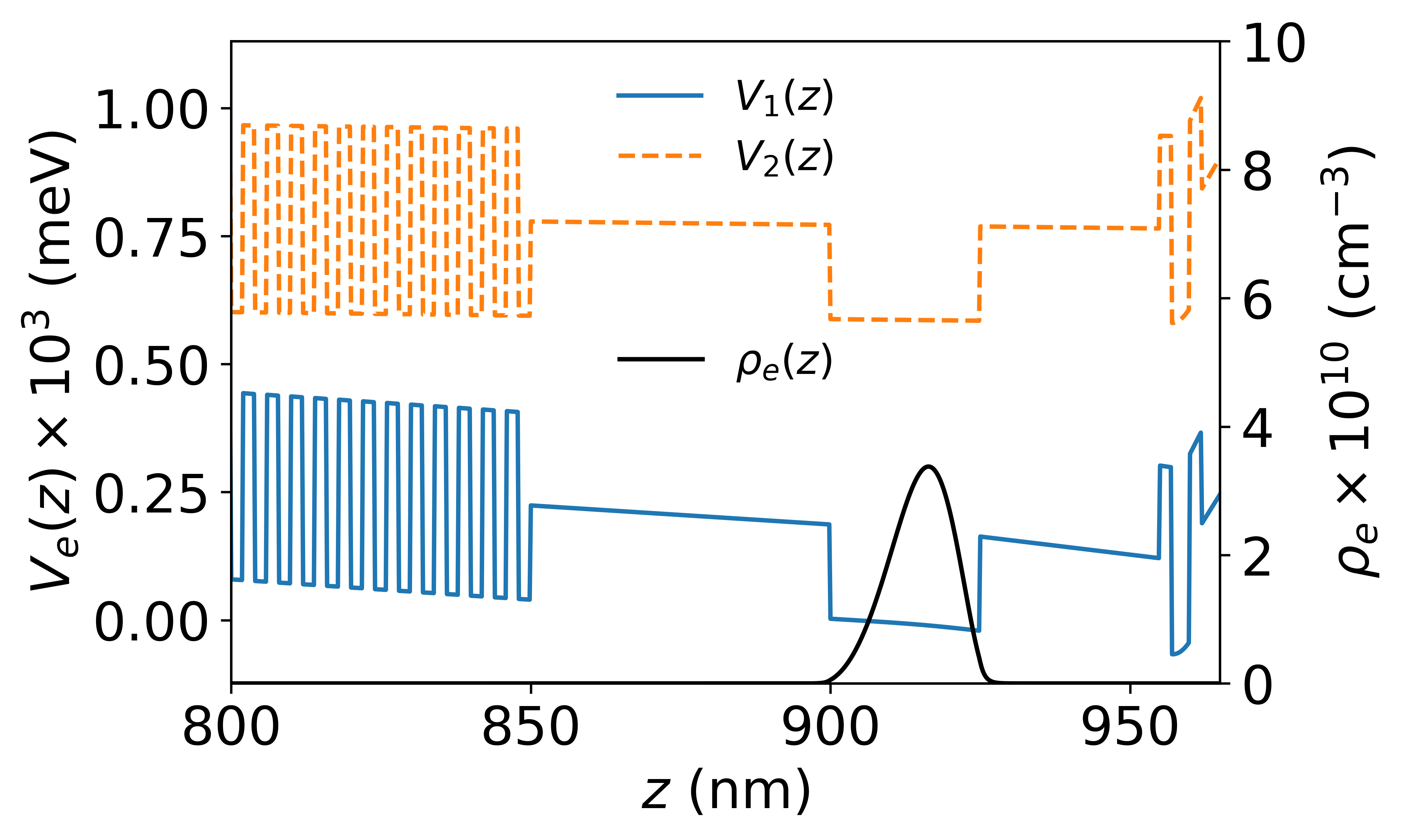}
    \caption{The potential profiles $V_1(z)$ and $V_2(z)$. Note, that $V_2(z)$ (etched case) is evaluated for Si doping  obtained for the unetched case. The solid black line indicates the carrier density $\rho_{e}(z)$ computed for potential $V_1(z)$, which corresponds to $n_{el}=\int_{z\in 2DEG}\rho_{e}(z)dz\approx 4.5\times 10^{10}$ cm$^{-2}$.}
\label{fig:zprofile}
\end{figure}
\footnote{Note that we excluded the bottom layer of $n$-doped AlGaAs as detailed information regarding the character of doping has not been provided ~\cite{DuLiu}}

\begin{figure}
\includegraphics[width=0.50\textwidth]{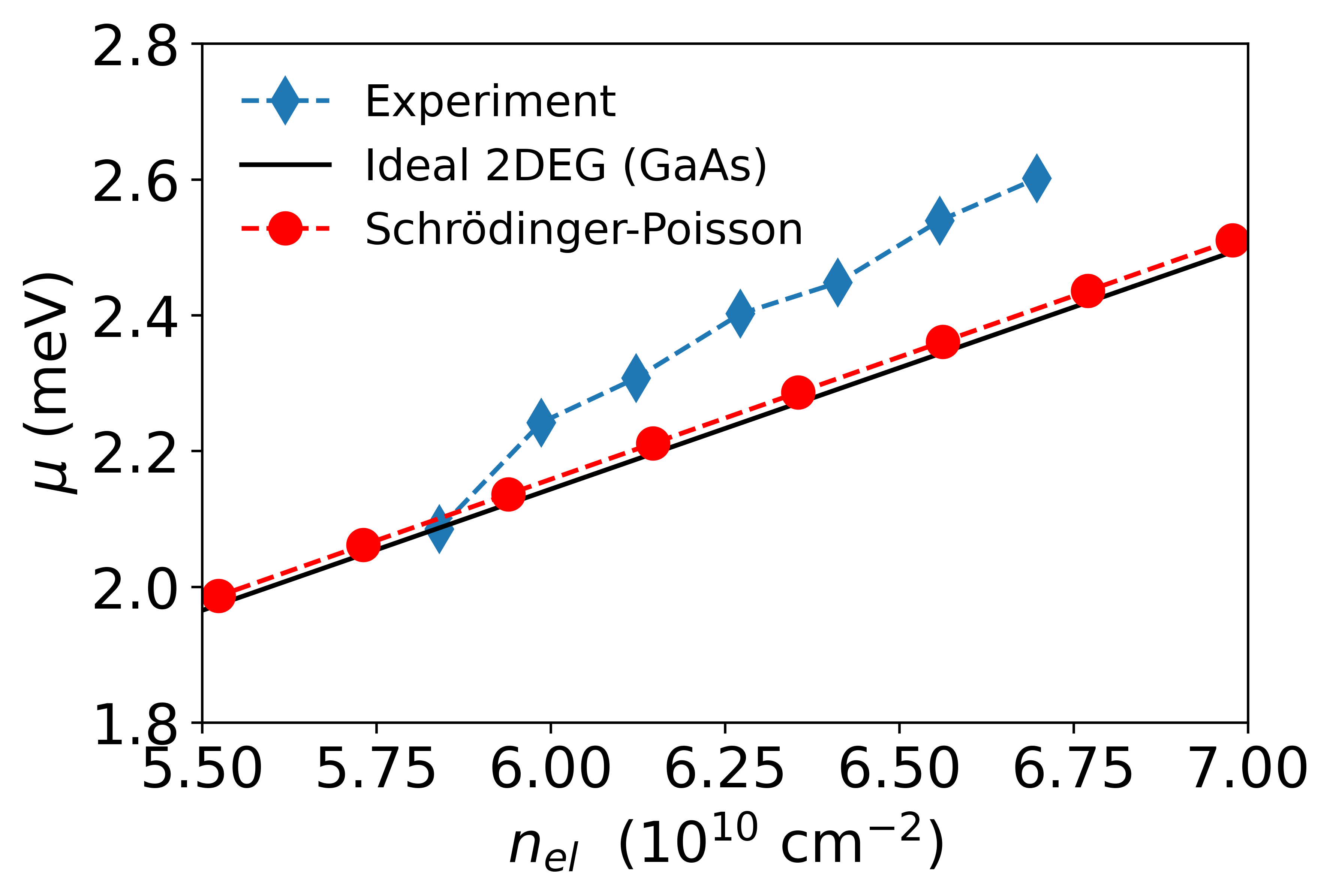}
\caption{The Fermi energy as a function of the electron concentration for the unpatterned 2DEG calculated within Schr\"{o}dinger-Poisson scheme and for the ideal 2DEG (red circles and black solid line, respectively) compared to the experimental data (blue diamonds).}
\label{fig:nvsexp}
\end{figure}

\begin{figure}
\includegraphics[width=0.50\textwidth]{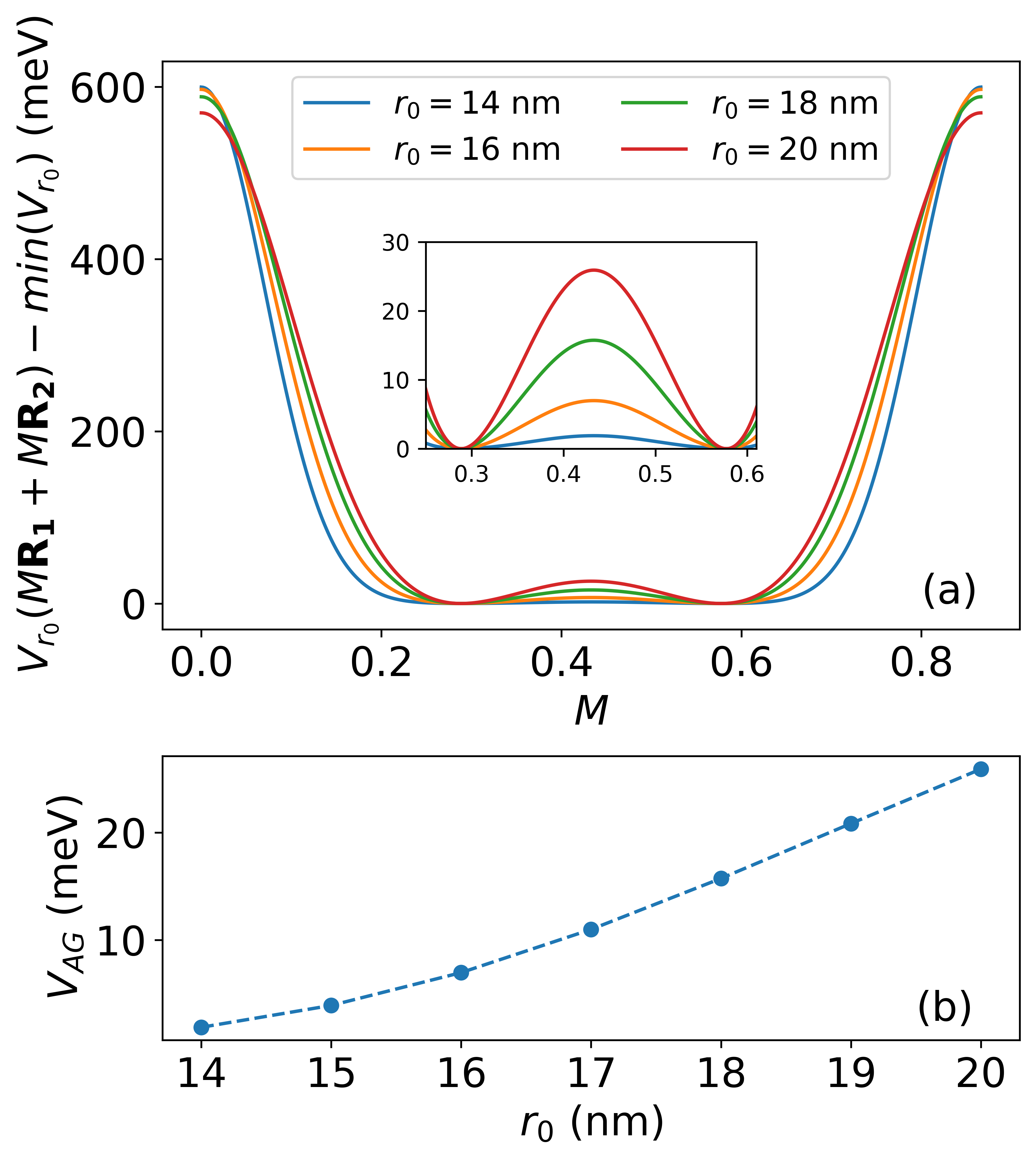}
\caption{(a) Cross section of the potential $V_{r_0}$ for different values of $r_0$. The increase in $r_0$ results in deeper local minima corresponding to the formation of a graphene-like lattice. (b) The depth of the local minima $V_{AG}$ as a function of $r_0$.}
\label{fig:potential}
\end{figure}

The resulting potential profiles $V_1(z)$ (non-etched case) and $V_2(z)$ (depth of etching $35 $ nm) are presented in Fig.~\ref{fig:zprofile}. The local maximum of the AD potential is estimated to be $V_0\approx600$ meV according to Eq.~\ref{eq:v0}. 
We confirm the reliability of our procedure by comparing the calculated Fermi energy $\mu$ as a function of $n_{el}$. In Fig.~\ref{fig:nvsexp} we present $\mu(n_{el})$ obtained for the potential profile $V_1(z)$, as well as for the available experimental data~\cite{DuLiu} and the dependence corresponding to the ideal 2DEG model. The Fermi energy for the considered concentration range is $2$ meV $\lesssim \mu \lesssim 2.5$ meV. Remarkably, the energies evaluated within Schr\"{o}dinger-Poisson scheme are in almost perfect agreement with those obtained for the ideal 2DEG with values only about $5\%$ lower than the experimental ones.
The estimated value of $V_0$ is finally applied to Eq.~\ref{eq:V}, which defines the planar 2D potential for a given parameter $r_0$. We focus on AD radii $15$ nm $\leq r_0\leq21$ nm. These values provide $V_{r_0}$ that exposes \emph{local} maxima at $\mathbf{R}_{lmax}=p/2 \times \mathbf{R}_1+q/2 \times \mathbf{R}_2$ and minima at $\mathbf{R}_{lmin}=u\times \mathbf{R}_1+v\times \mathbf{R}_2$ where $p,q\in \mathbb{Z}$ and $u,v\in\{p/3,2q/3\}$ (Fig.~\ref{fig:potential}(a)). In particular, the points defined by the vectors $\mathbf{R}_{lmin}^{uv}$ correspond to the honeycomb lattice. The depth of the local minima $V_{AG}=V_{r_0}(\mathbf{R}_{lmax})-V_{r_0}(\mathbf{R}_{lmin})$  depends on $r_0$. For the lower range of $r_0$ considered, $V_{AG}$ is only of the order of a few microelectronvolts, while for $r_0=20$ nm $V_{AG}\approx26$ meV, as derived from the data shown in Fig.~\ref{fig:potential}(b). Thus, it may be expected that a sufficiently large $r_0$ enforces a stronger confinement of electrons and, in turn, their localization in honeycomb trapping centers, which gives the opportunity for emerging strongly correlated state. 

\subsection{Artificial lattice: Mean-field approach}
The elaboration of $V_{r_0}(z)$ eventually allows us to perform an analysis of the electronic properties of the formed artificial honeycomb lattice. First, we provide a mean-field approach in which we decouple the modulated 2DEG from the rest of the system. It is done by factorizing a single-electron wave function $\Psi_n(\mathbf{r},z)=\psi_n(\mathbf{r})\phi_1(z)$.
%where $\phi(z)\approx\sqrt{\frac{2}{d}}\sin{\big(z\frac{\pi}{d}\big)}$ can be aproximated by a solution for an infinite rectangular quantum well of width $d$. 
Within this assumption, to estimate whether the distribution of electronic density (at a concentration of two electrons per elementary cell) for the estimated potential is honeycomb-like, we solve again the Schr\"odinger-Poisson problem,this time, in $x-y$ plane for $\psi_n(\mathbf{r})$, i.e.,
\begin{subequations}
\label{eq:2d_set}
\begin{gather}
    \Big[-\frac{\hbar^2}{2m^{*}_{||}}\nabla^2_{r}+V_{r_0}(\mathbf{r})+V_c(\mathbf{r})\Big]\psi_n(\mathbf{r})=E_{n}\psi(\mathbf{r})\\
    \nabla^{1}_r V_c(\mathbf{r})=-\frac{\rho_r(\mathbf{r})}{\epsilon_0\epsilon_r}\\
    \rho_r(\mathbf{r})=\frac{1}{d}\sum_{n,\sigma}^{N_{el}/2}|\psi_{n\sigma}(\mathbf{r})|^2,
\end{gather}
\end{subequations}
where $N_{el}=2$ is the number of electrons in the unit cell and $\sigma=\{\uparrow,\downarrow\}$ stands for the $S^{z}$ spin component. The factor $1/d$ in Eq.~\ref{eq:2d_set}c refers to \emph{mean} electron concentration (per spin) with respect to $z$ direction in 2DEG. Eventually  the estimated \emph{total} mean field 2D potential at given $r_0$ is modeled as a sum of $V_{r_0}(\mathbf{r})$ and $V_c(\mathbf{r})$ resulting from the self-consistent numerical solution of Eqs.~\ref{eq:2d_set}(a-c), that is,
\begin{align}
\Tilde{V}(\mathbf{r})\equiv V_{r_0}(\mathbf{r})+ V_c(\mathbf{r}).
\end{align}

Eventually, $V_{r_0}$ is used to solve the two-dimensional mean field model defined in Eqs.~(\ref{eq:2d_set}a-c). In our calculations we take a single unit cell and assume periodic boundary conditions. A computational procedure is performed on the triangular mesh consisting of $70\times70$ nodes. We consider the half-filled case, that is, $\int_{\Omega}\rho_{r}(\mathbf{r})d\mathbf{r}=2$, where $\Omega$ is an area of the unit cell. 

\begin{figure}
\includegraphics[width=0.50\textwidth]{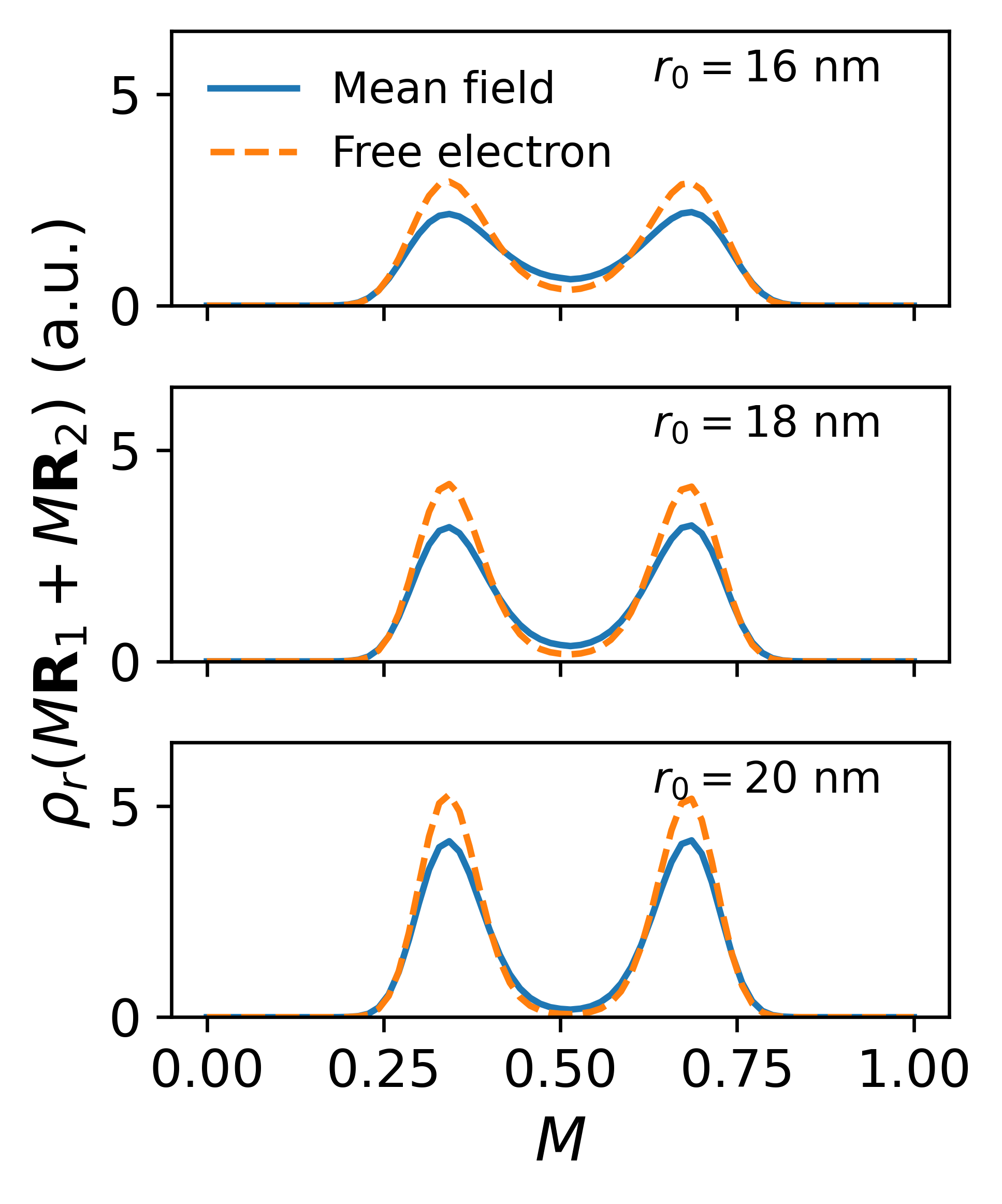}
\caption{Cross sections of the resulting electronic density along $\mathbf{R}_1+\mathbf{R}_2$ vector  for the three representative values of $r_0$.}
\label{fig:poisson2d}
\end{figure}

This self-consistent mean-field (MF) approach reveals a meaningful renormalization of the density distribution compared to that of the free electron (non-interacting) case. Namely, the inclusion of Coulomb interactions leads to the more uniform \emph{smearing} of carriers throughout $\Omega$ as can be deduced from Fig.~\ref{fig:poisson2d}. Particularly, this tendency is meaningful  for the lower values of $r_0$, as $V_{AG}$ increases with radius of AD.

\section{Wannier basis and Single-particle Hamiltonian}

The task of constructing the Wannier basis is crucial in view of the reliable determination of both the single-particle Hamiltonian in Tight Binding Approximation (TBA) and the electron-electron interaction amplitudes. In the considered system, the minima of the trapping centers are expected to be shallow, specifically for the semimetallic case, and the distance between the neighboring dots in the honeycomb lattice is $\sim r_0$. Here, the Wannier functions are obtained by application of the \emph{Projection Method}~\cite{Marzari} adapted to a 2D case. This method is based on projecting smooth trial functions $g_n(\mathbf{r})$ centered at positions that are supposed to correspond to the maxima of the electronic density onto Bloch states represented by $ \psi_{n,\mathbf{k}}(\mathbf{r})$ (where $n$ refers to the band index). Subsequently, the procedure that provides unitary transformation is carried out, which eventually results in well-localized Wannier states when $\{g_n(\mathbf{r})\}$ are correctly guessed. Namely, we obtain $\psi_{n,\mathbf{k}}(\mathbf{r})$ by numerical diagonalization of the single-particle mean-field Hamiltonian $\mathcal{\hat{H}}_0$ defined as
\begin{align}
    \mathcal{\hat{H}}_0=-\frac{\hbar^2\mathbf{k}^2}{2m^{*}_{||}} + \Tilde{V}(\mathbf{r}).
\label{eq:h0}
\end{align}

Next, we take the auxiliary basis $\{\ket{\phi_{n\mathbf{k}}}\}$ defined as
\begin{align}
\ket{\phi_{n\mathbf{k}}}=\sum_m \ket{\psi_{m\mathbf{k}}}\braket{\psi_{m\mathbf{k}}|g_n},
\label{eq:expansion}
\end{align}
where the summation index $m$ enumerates the considered bands, to construct the L\"{o}wdin-orthonormalized Bloch-like states~\cite{Marzari}
\begin{align}
    \ket{\Tilde{\psi}_{n\mathbf{k}}}=\sum_m \ket{\phi_{m\mathbf{k}}}\big(S_{\mathbf{k}}^{-1/2}\big)_{mn},
\end{align}
where $\big(S_{\mathbf{k}}\big)_{mn}=\braket{\phi_{m\mathbf{k}}|\phi_{n\mathbf{k}}}$.
Eventually, Wannier states $\{w_{n\mathbf{R}_{ij}}\}$ are obtained by applying the standard Fourier transform, i.e.,
\begin{align}
    \ket{w_{n\mathbf{R}_{ij}}}=\frac{1}{2}\frac{L^2\sqrt{3}}{(2\pi)^2}\int_{BZ}d\mathbf{k}\exp\big({-i\mathbf{k}\cdot\mathbf{R_{ij}}}\big)\ket{\Tilde{\psi}_{n\mathbf{k}}}.
\end{align}
 
For each considered $r_0$, the numerical representation of the Hamiltonian given in Eq.~\ref{eq:h0} is implemented in KWANT package~\cite{Kwant} and diagonalized in the momentum space for the particular wave vector $\mathbf{k}$. We probe $32\times32$ equally spaced points from the first Brillouin in the unit cell spanned by vectors $\mathbf{G}_1=\frac{2\pi}{L}\Big(1,-\frac{\sqrt{3}}{3}\Big)$ and $\mathbf{G}_2=\frac{2\pi}{L}\Big(0,\frac{2\sqrt{3}}{3}\Big)$ that conform to the standard relation $\mathbf{G}_i\cdot \mathbf{R}_j=2\pi \delta_{ij}$. 
We choose Gaussian trial functions $g_n(\mathbf{r})$,
\begin{align}
 g_n(\mathbf{r})\equiv\frac{1}{2\pi\sigma^2}\exp{\Bigg[-\Bigg(\frac{|\mathbf{r}-\mathbf{R}_{lmin}^{n}|}{\sigma}\Bigg)^2\Bigg]}
\end{align}
with $n\in\{\alpha,\beta\}$, where $\alpha$ and $\beta$ correspond to two minima present in the confining potential surface at $\mathbf{R}^{\alpha}_{lmin}=L(1/2,\sqrt{3}/6)$ and $\mathbf{R}^{\beta}_{lmin}=L(1,\sqrt{3}/3)$ respectively. Thus, we construct the Wannier basis properly describing the two lowest bands of the system. The presented results correspond to $\sigma=40$ nm, although our test calculations have not revealed any significant differences for $\sigma$, which are $20$ and $60$ nm. The integration manifold in Eq.~\ref{eq:expansion} has been chosen to be the area corresponding to $5\times5$ unit cells to ensure the proper decay of $g_n(\mathbf{r})$ for the assumed value of $\sigma$.

\begin{figure}
\includegraphics[width=0.5\textwidth]{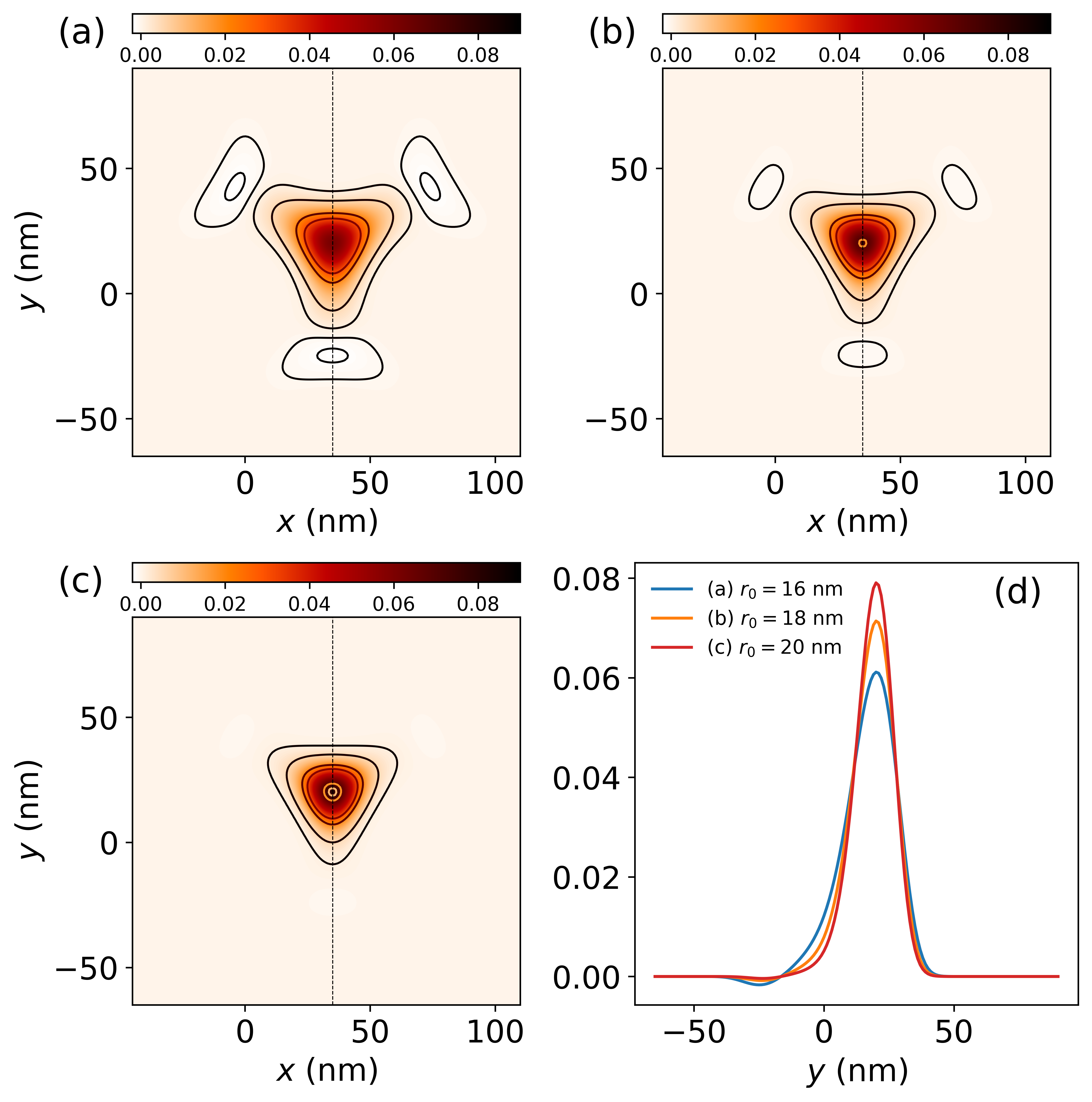}
\caption{2D plot of Wannier functions $w_{\alpha R_{00}}(x,y)$ for $r_0=16$ nm (a), $r_0=18$ nm (b) and $r_0=20$ nm (c). In (d) their cross sections along $x=35$ nm (dashed vertical lines in (a-c)) are presented as a function of $y$. The disappearance of local minima with an increasing value of $r_0$ can be identified in terms of isoline inspection (a-c), which is accompanied by an increase in maximal amplitude at the localization center (a-d).}
\label{fig:wannier2d}
\end{figure}

In Fig.~\ref{fig:wannier2d}(a-c) we show representative plots of the Wannier functions for $n=\alpha$ and different values of the parameter $r_0$. Note that they preserve the three-fold rotational symmetry $\mathbf{C_3}$ as desired. The same feature holds for $w_{\beta R_{ij}}(\mathbf{r})$ since it is reproducable from $w_{\alpha R_{ij}}(\mathbf{r})$ by means of the following transformation,
\begin{align}
w_{\beta R_{lm}}(\mathbf{r}) = \mathbf{\hat{T}}_{\mathbf{R}_{l-i,m-j}}\mathbf{\hat{O}}_{\mathbf{R}_{ij}+\mathbf{R}_2-\mathbf{R}_1} w_{\alpha R_{ij}}(\mathbf{r}),
\label{eq:TO}
\end{align}
where $\mathbf{\hat{T}}$ is the translation by the vector $\mathbf{R}_{l-i,m-j}$ and $\mathbf{\hat{O}}$ is  operator of reflection with respect to the direction given by ${\mathbf{R}_{ij}+\mathbf{R}_2-\mathbf{R}_1}$. Therefore, we disregard the analysis for the case $n=\beta$ as the one with $n=\alpha$ is fully representative; however, we emphasize that Eq.~\ref{eq:TO} holds (within numerical precision) also from the perspective of our numerical calculations. The validity of the procedure has been confirmed by inspecting the orthogonality, which, by construction, should be fulfilled exactly. We have found that this requirement is met within the numerical precision. 

From the data contained in Fig.~\ref{fig:wannier2d} one may conclude that the spatial extent of the resulting Wannier states shrinks with increasing $r_0$, which can be regarded as a natural consequence of mutual enforcement of trapping. However, since these states are obtained with electron-electron interactions at the level of the mean-field approach, we do not find it trivial, since the confining role of the local minima in $V_{r_0}$ is compatible with respect to $V_c$ as comes from Fig.~\ref{fig:poisson2d}. Furthermore, in addition to enhanced localization, the nodal lines that reside in the vicinity of the location of the nearest neighbor (nn) centers and the three local minima enclosed by them disappear for $r_0$ approaching $20$ nm. 

In the next step, evaluated Wannier functions are subsequently utilized for the construction of TBA Hamiltonian given explicitely in second quantization fromalism as
\begin{align}
\mathcal{\hat{H}}_{TB}=\sum_{\mu\nu,\sigma}t_{\mu\nu}\hat{c}_{\mu,\sigma}^{\dagger}\hat{c}_{\nu,\sigma}^{},
\label{eq:TBA_Hamiltonian}
\end{align}
The $\mu,\nu\in\{(i,j),n\}$ in Eq.~\ref{eq:TBA_Hamiltonian} are defined by relation $\mathbf{R_{\mu}}\equiv \mathbf{R}_{ij}$, where $n$  number kind of Wannier function, $\hat{c}_{\mu,\sigma}^{\dagger}$($\hat{c}_{\mu,\sigma}^{}$) creates (annihilates) an electron of spin  $\sigma$  associated with appropriate $w_{\mu}(\mathbf{r})$, and,
\begin{align}
    t_{\mu\nu}\equiv\Big\langle w_{\mu}(\mathbf{r})\Big|-\frac{\hbar^2}{2m^{*}_{||}}\nabla^{2}_{\mathbf{r}}+\Tilde{V}(\mathbf{r})\Big|w_{\nu}(\mathbf{r})\Big\rangle.
\label{eq:spintegrals}
\end{align}

For the sake of clarity, we introduce the following notation,
\begin{align}
    t_{\mu\nu}\equiv t_{I(\mu,\nu)}:I(\mu,\nu)= Z \in \{0,1,2,3,..\},
    \label{eq:tdefinition}
\end{align}
where $I(\mu,\nu)$ maps the pair of position-orbital indices into the proper natural number $Z$ labeling $Z$-th as the nearest neighbor corresponding to the pair $(\mu,\nu)$.  

The notation is completed by assigning
\begin{align}
    t_{\mu\mu}=t_0\equiv \epsilon_0,
\end{align}
as is the convention commonly used.

\begin{figure}
\includegraphics[width=0.5\textwidth]{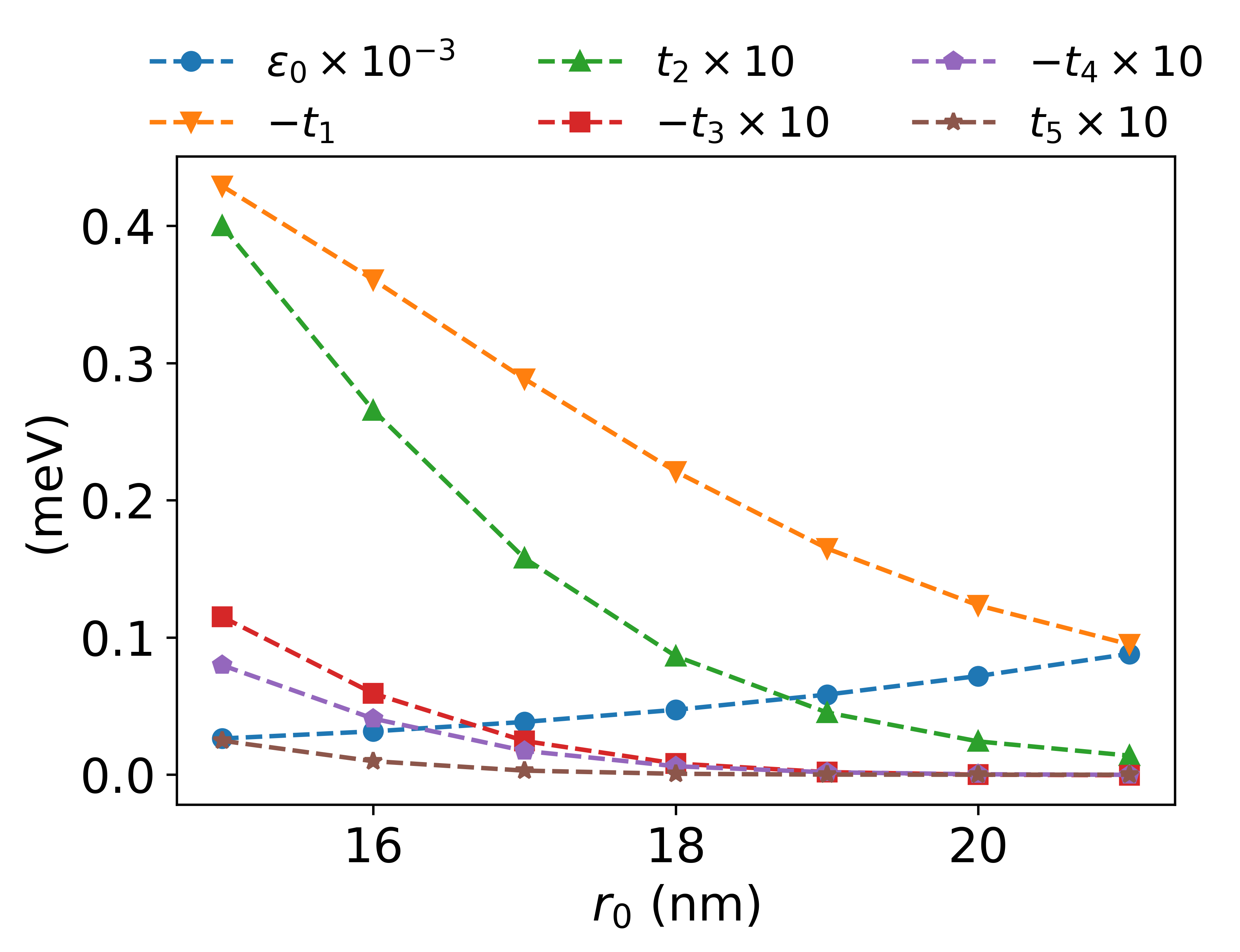}
\caption{The single particle amplitudes as a function of $r_0$ up to the $5$th-nearest neighbor. The hoppings referring to $Z\geq5$ play a marginal role in view of the resulting band structure for each of the examined $r_0$.}
\label{fig:hoppings}
\end{figure}

In Fig.~\ref{fig:hoppings} we present values of single particle amplitudes up to $Z=5$ as a function of $r_0$. When the size of the trapping center shrinks, the value of $\epsilon_0$ increases. The opposite holds for $|t_1|$, which is, however, of the order of $10^{-1}$ meV in the entire examined range of $r_0$. The hopping values $t_2$ are positive and of an order of magnitude lower than $|t_1|$. Note that we have computed the hopping integrals up to $Z=17$ and found their rapid decay with increasing $Z$, that is, $|t_{Z\geq5}|\lesssim 10^{-3}$ meV for each value of $r_0$.

\begin{figure}
\includegraphics[width=0.5\textwidth]{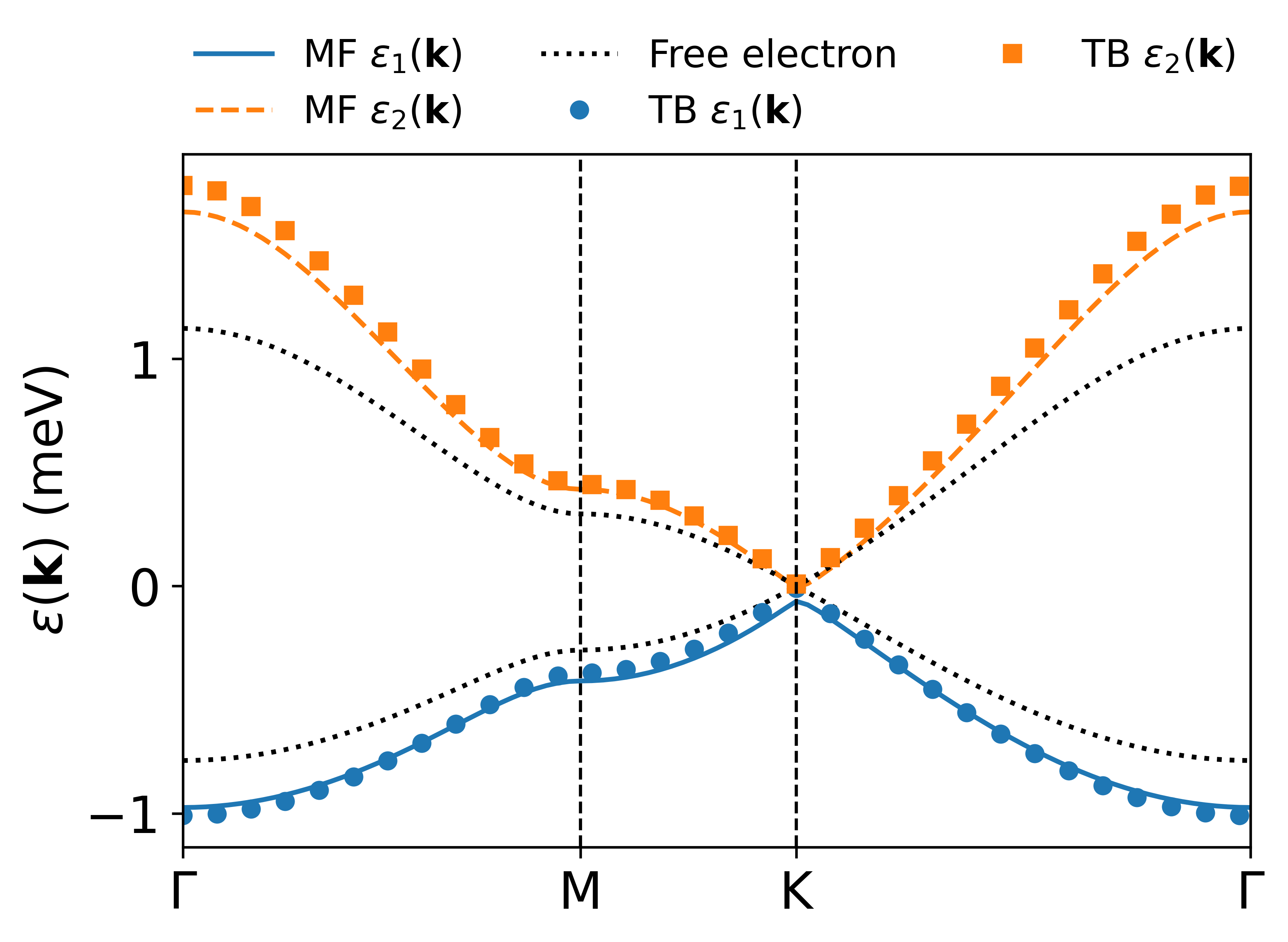}
\caption{Dispersion relations for the two lowest bands obtained for $r_0=15$ nm. MF and TB refer to $\mathcal{\hat{H}}_{0}$ and $\mathcal{\hat{H}}_{TB}$ respectively. The values of $\epsilon_{1,2}(\mathbf{k})$ resulting from the diagonalization of $\mathcal{\hat{H}}_{TB}$ in a momentum space are in good agreement with those that come from the solution of the eigenproblem for $\mathcal{\hat{H}}_{0}$. Bands represented by dotted black lines correspond to the free-electron approach.} 
\label{fig:dispersion}
\end{figure}

Finally, the obtained sets $\{t_i\}_{r_0}$ are utilized to formulate $\mathcal{\hat{H}}_{TB}$ for each value of $r_0$. In Fig.~\ref{fig:dispersion} we present a spectrum for $r_0=15$ nm that gives $\epsilon{1,2}(\mathbf{k})$  which is in good agreement with experimental evidence ~\cite{DuLiu}. That is, the formation of the \emph{Dirac cone} in the vicinity of $\mathbf{K}$ can be clearly identified. Notably, the difference $\Delta\varepsilon_M\equiv\varepsilon_2(\mathbf{M})-\varepsilon_1(\mathbf{M})$ is of particular interest, since as $\mathbf{M}=(\pi,\pi/\sqrt{3} )/L$ it can be confronted against the energies of van Hove singularity peaks in the photolouminescence spectra. As comes from Fig.~\ref{fig:dispersion}, $\Delta\varepsilon_{M}\approx1$ meV is close to the experimental value  $\sim0.9$ meV. Namely, $\varepsilon_{1,2}$ resulting from both approaches are nearly identical, which validates the procedure covering Wannier basis elaboration, as well as the subsequent calculation of hopping parameters presented in the further  part of the paper.

The dispersion relation obtained by the free electron approach provides a significantly different $\Delta\varepsilon_M$, which is $\sim0.6$ meV. Also, in this case, both bands are narrower compared to the MF approach. We do not find perfect agreement of $\Delta\varepsilon_{M-\Gamma}\equiv\varepsilon_1(\mathbf{M})-\varepsilon_1(\mathbf{\Gamma})\approx0.6$ meV with respect to the value $\approx 1$ meV provided by Du et al. However, from the data contained in their work it is difficult to extract the precise value of $\varepsilon(\mathbf{\Gamma})$ for carrier concentrations that result in the presence of ALD. Also, since they indicate that the emergence of the honeycomb lattice results from the filling of the lower band, resulting in the occupation of states for which $\varepsilon({\mathbf{k_F}})$ is slightly below $\varepsilon({\mathbf{M}})$, we consider this disagreement to be a minor issue.

Importantly, with increasing $r_0$ we observe a decreasing value of $\Delta\varepsilon_M$, as well as a narrowing of the entire band structure, reflecting a stronger confinement of electrons in trapping centers. Therefore, it may be expected that at some $r_0$ the enhancement of carrier localization causes electron-electron interactions to dominate over kinetic energy and, consequently, drives the system to the strongly correlated regime. To clarify this presumption, we further study the interaction amplitudes, construct a full electronic Hamiltonian, and find its approximate ground state by means of VMC method for the selected sets of interactions.

\section{Interacting picture}
\subsection{Electronic interactions and Hamiltonian}
The determination of Wannier functions and $\mathcal{\hat{H}}_{TBA}$ allows to evaluate the complete electronic Hamiltonian $\mathcal{\hat{H}}$,
\begin{align}
    \mathcal{\hat{H}}=\mathcal{\hat{H}}_{TB} + \mathcal{\hat{H}}_{e-e},
\end{align}
where  $\mathcal{\hat{H}}_{e-e}$ contains electron-electron interaction terms. 
Although the finite width of 2DEG affects $\mathcal{\hat{H}}_{TB}$ only by the additive constant (related to confinement in the $z$ direction), it is not the case for the interacting part~\cite{Biborski0}. Thus, $\mathcal{\hat{H}}_{e-e}$ takes the following form
\begin{subequations}
\begin{gather}
\mathcal{\hat{H}}_{e-e}=\frac{1}{2}\sum_{\substack{\mu,\nu,\\ \gamma,\tau}}
\sum_{\sigma,\sigma'}V_{\mu\nu\gamma\tau}\hat{c}^{\dagger}_{\mu,\sigma}\hat{c}^{\dagger}_{\nu,\sigma'}\hat{c}_{l,\tau,\sigma'}\hat{c}_{\gamma,\sigma},\\
V_{\mu\nu\gamma\tau}= \Big\langle \widetilde{w}_{\mu}(\mathbf{r},z)\widetilde{w}_{\nu}(\mathbf{r'},z')\Big|\hat{V}_{e-e}\Big|\widetilde{w}_{\gamma}(\mathbf{r},z)\widetilde{w}_{\tau}(\mathbf{r'},z')\Big\rangle,\\
\widetilde{w}_{\mu}(\mathbf{r},z)\equiv w_{\mu}(\mathbf{r})\phi(z).
\end{gather}
\label{eq:interactiontensor}
\end{subequations}

Note, that we take $\phi(z)=\phi_1(z)$, i.e., the lowest energy state obtained within Schrödinger-Poisson scheme as described in Sec. II A.
The electron-electron interaction operator $\hat{V}_{e-e}$ is taken in the Yukawa potential form, to account for the screening effects among electrons in 2DEG (see Section Results), that is,
\begin{align}
    \hat{V}_{e-e}=\frac{e^2\exp{\Big(-q_{TF}\sqrt{|\mathbf{r}-\mathbf{r}'|^2+(z-z')^2}\Big)}}{4\pi\epsilon_0\epsilon_r \sqrt{|\mathbf{r}-\mathbf{r}'|^2+(z-z')^2}},
\label{eq:veeoperator}
\end{align}
where $e$ is the electron charge and $q_{TF}$ is the estimated length of the Thomas-Fermi wave vector.

Evaluation of microscopic parameters ($t_{\mu\nu}$,$V_{\mu\nu\gamma\tau}$) eventually leads to formulation of the Hamiltonian $\mathcal{\hat{H}}$, which can be diagonalized only by approximate methods. Here, we exploit the Variational Monte Carlo technique~\cite{Biborski0} as will be presented at the end of this section.

The leading elements of the electron-electron interaction tensor $V_{\mu\nu\gamma\tau}$ provided in Eq.~\ref{eq:interactiontensor} can be obtained numerically. However, to perform this task, the length of the Thomas-Fermi wave vector $q_{TF}$ in Eq.~\ref{eq:veeoperator}---which tunes the magnitude of screening---needs to be reasonably estimated. 

The screening, directly related to the density of the electron gas, together with the local confinement of the carriers, may play a key role in the tuning of the magnitude of electron-electron interactions~\cite{Vandersypen}. In our case, the Wigner-Seitz radius $r_s^{nD}$ (where $D$ stands for dimensionality) at $n_{el}\approx 4.5\times10^{10} $ cm$^{-2}$ is $r_s^{2D}\approx 2.66$ when treating the system as strictly two-dimensional, and $r_s^{3D}\approx2.36$ when one considers it as quasi-two-dimensional, that is, as being a layer of width $d=25$ nm. Here, both values of $r_s$ are accidentally close to each other. As they are $\sim2.5$, one concludes that electron-electron interactions and kinetic energy are of similar magnitude~\cite{Kotov,Ortiz}. Thus, disregarding the modulation described by $V_{r_0}$ for a while, the system may be safely considered to be the one in which the influence of Coulomb interactions between carriers is moderate with respect to its electronic properties, namely $1\lesssim r_s\lesssim10$. The system falls into the class of the so-called \emph{intermediate regime}~\cite{Ortiz} with respect to the relation between electron-electron interactions and their kinetic energy. 

Also, as for $r_0=15$ nm $V_{r_0}$ exhibits only shallow local minima and the resulting density profiles are characterized by the nearly homogeneous distribution of electrons (excluding areas of ADs) with relatively weak local maxima located at honeycomb lattice sites, it seems reasonable to assume metallic-like screening for this case. Therefore, increasing $r_0$ should also affect the screening length. However, assuming its value as that in the non- or weakly confinig regime, one only overestimates the magnitude of $q_{TF}$ and when interactions become significantly greater than hopping amplitudes, one may deduce that the strongly correlated regime is achieved. Therefore, such a strategy provides only an overestimation of the minimal values of $r_0$ for which ALD should exhibit typical phenomena for the strongly correlated system.

\begin{figure}
\includegraphics[width=0.5\textwidth]{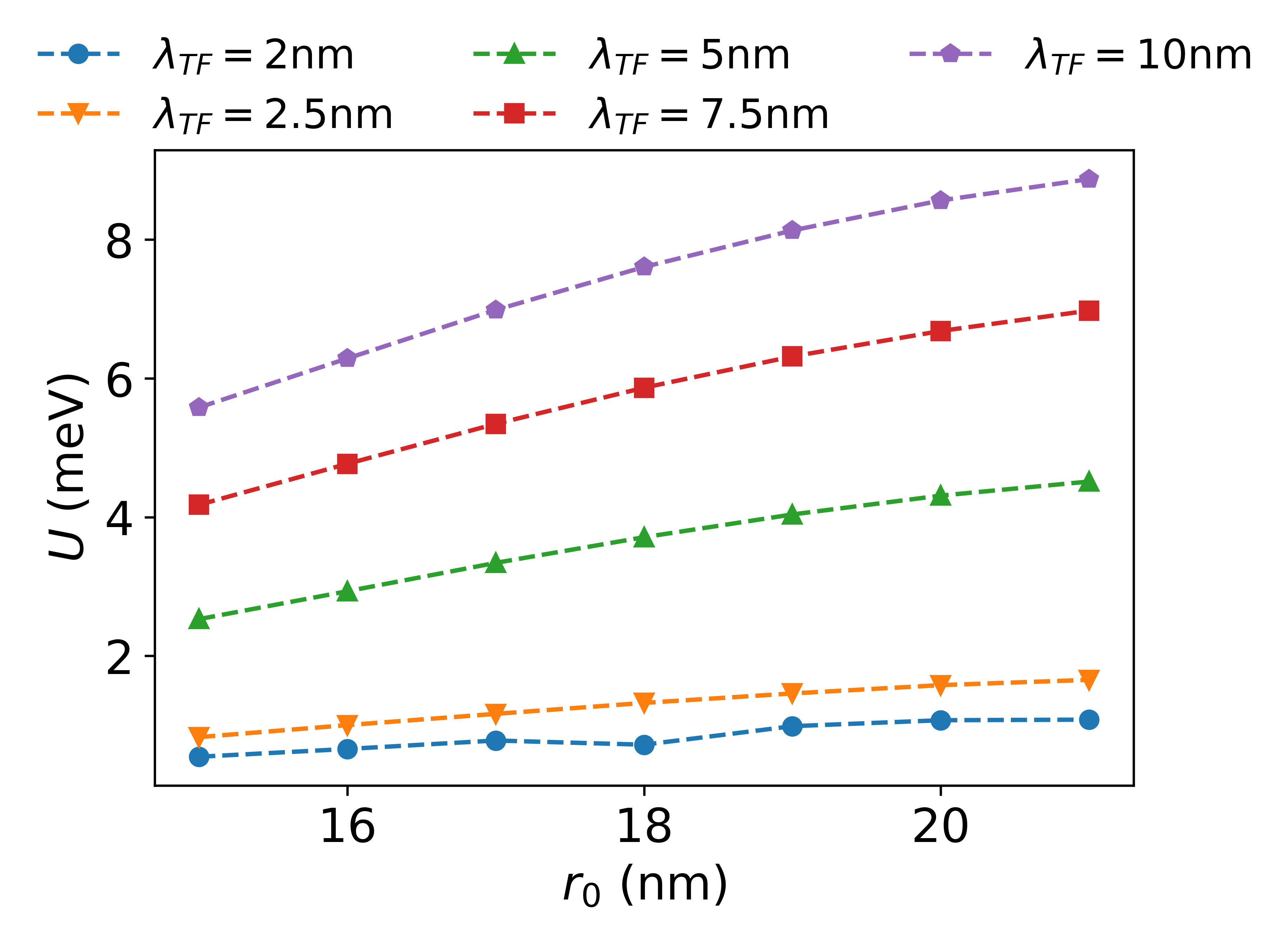}
\caption{The amplitudes of Hubbard $U$ on-site interaction  as a function of $r_0$ for the selected values of $\lambda_{TF}$.} 
\label{fig:Uvalues}
\end{figure}

We inspect the values of $1/q_{TF}\equiv\lambda_{TF}\in[2\text{ nm},10 \text{nm}]$---where $\lambda_{TF}$ is the Thomas--Fermi length, which is supported by the following estimations. First, we may consider the  $q_{TF}$ resulting  from the ideal 2DEG approach assuming $m^{*}=0.067m_{el}$ and $\epsilon_r\approx13$, that is, the specific values of the effective mass and dielectric constant of the GaAs material, respectively. In this case, since $q_{TF}$ does not depend on the carrier concentration, $\lambda_{TF}=a_{B}^{*}/2\approx5$ nm, where $a_B^{*}$ is the effective Bohr radius. On the other hand, taking the 3D approach and $n_{el}^{3D}=4.5\times10^{10}/d\text{ cm}^{-2}$, one is left with $\lambda_{TF}\approx9$ nm, thus a value nearly twice higher than for the 2D case. Eventually, $\lambda_{TF}$ can be estimated on the basis of a quasi-2D model (slab) of 2DEG elaborated by Moreno and M\'endez-Moreno~\cite{Moreno}. Their approach leads to $\lambda_{TF}\approx2$ nm for the slab width of $d=25$ nm, i.e. a value substantially lower than that of the ideal two-dimensional electron gas. As the formulation of a general description of screening and its role in a quasi-two-dimensional system is a complex problem that is beyond the scope of this work, we examine the values of $\lambda_{TF}$, relying on the estimations presented above.

\begin{figure}
\includegraphics[width=0.45\textwidth]{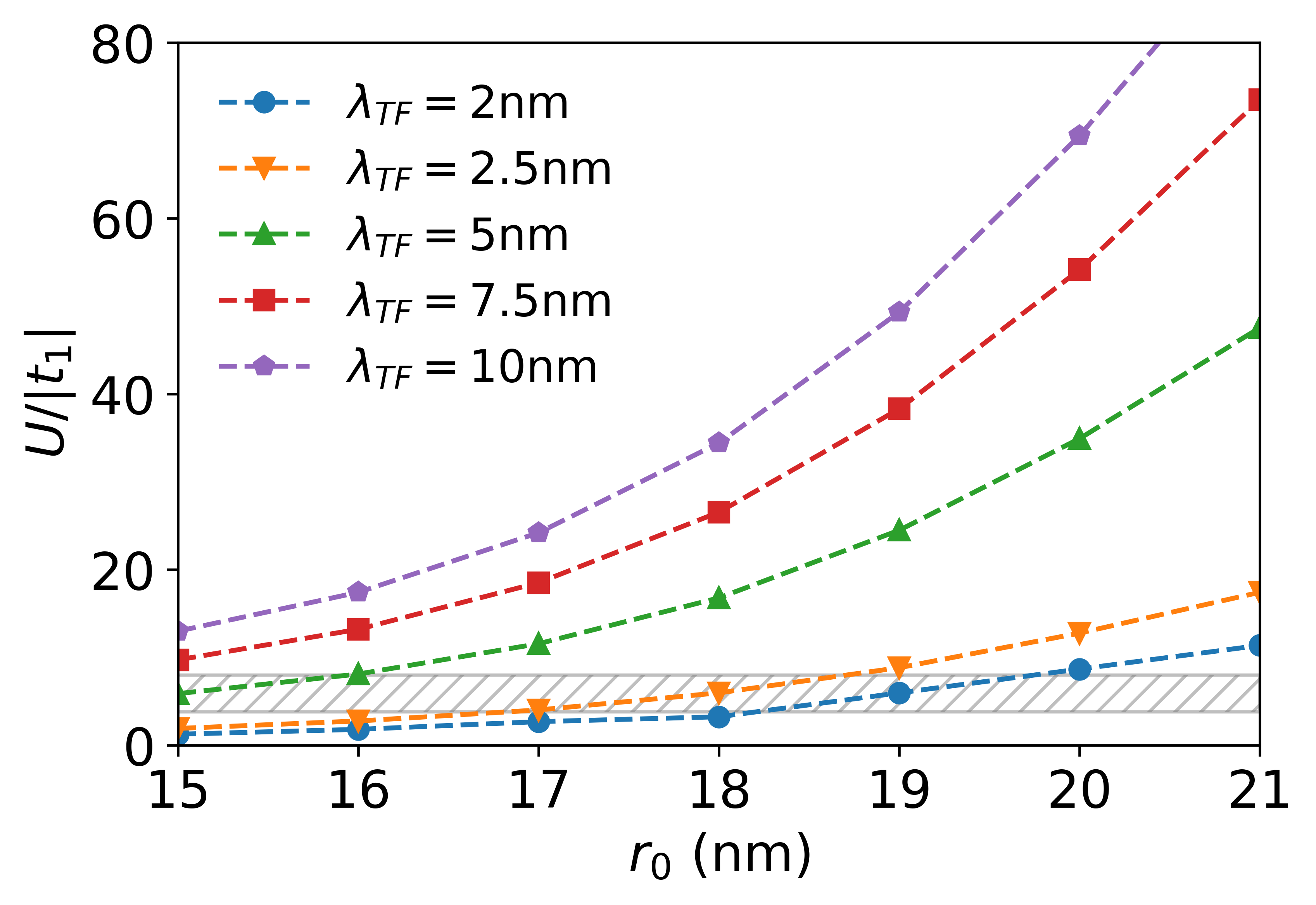}
\caption{The ratio between Hubbard on-site repulsion $U$ and the absolute value of $t_1$. The dashed area corresponds to $U_c/|t_1|\approx3.8\div4.2$ for which the AFMI phase is supposed to emerge when considering the Hubbard model on the honeycomb lattice~\cite{Sorella2012,Meng2010}.} 
\label{fig:Uvst}
\end{figure}

We compute interaction amplitudes of the form $V_{\mu\nu\mu\nu}$, that is, we take into account the on-site Hubbard interaction $U\equiv V_{0,0}$ (applying the same rule of indexing as for hoppings---see Eq.~\ref{eq:tdefinition}), as well as intersite interactions $K_{0,Z}$ with $Z$ up to $5$. The calculation procedure has been carried out using the Cuba library~\cite{Cuba, Biborski0} designed for the integration of multi-variable functions. In Fig.~\ref{fig:Uvalues} we present values of $U$ as a function of radius $r_0$ for the selected screening lengths $\lambda_{TF}$. 
%The estimated numerical error does not exceed $0.1$ meV. 
According to the increasing localization of Wannier functions with $r_0$, the on-site repulsion amplitudes also increase. For example, for $\lambda_{TF}=2.5$ nm and $r_0=15$ nm $U\approx0.6$ meV, while for the same screening length but for $r_0=21$ nm, we obtain $U\approx1.3$ meV.
Importantly, for $\lambda_{TF}\gtrsim 5$ nm, the ratio $U/|t_1|\gtrsim3.8$ for $r_0=15$ nm, as can be seen in Fig.~\ref{fig:Uvst}, which may suggest that the system is in the vicinity of the transition point between the semi-metallic and antiferromagnetic Mott insulator (AFMI) phases~\cite{Sorella2012}. However, since we identify the radius $r_0=15$ as the one that corresponds to the experimental setup for which there is no evidence of AFMI formation~\cite{DuLiu}, we conclude that $\lambda_{TF}$ should be considered as $\lesssim5$ nm in the framework of the elaborated model. 

Note that since $|t_1|$ is one order of magnitude greater than $t_2$ for the entire range of $r_0$, the estimate of critical radius  $r_0^c$ for which AFMI possibly emerges by inspecting $U/|t_1|$ is justified, since in this view the electronic properties of the system should be similar to those of the Hubbard model on the honeycomb lattice. Thus, taking $\lambda_{TF}=2$ nm, we find that the transition should occur at $r_0^c\lesssim20$ nm.

Next, we examine the intersite amplitudes $K_{0,Z}$. We find that for $\lambda_{TF}\leq5$ nm, they are at least two orders of magnitude smaller than the corresponding $U$ for each considered $r_0$. Therefore, the interplay between $U-K$ is unlike to cause the formation of a charge density wave (CDW), which is believed to compete with the semimetallic and AFMI phases when the magnitude of intersite interactions is significant~\cite{WuWei}. Eventually, according to the values of the dominating interaction amplitudes, we expect a transition between the semi-metallic and AFMI phases when the value of $r_0$ increases. We illustrate this trend by the results obtained from our VMC calculations for $\lambda=2.5$ nm as presented below.

\subsection{Ground state of interacting Hamiltonian by means of VMC method}
We formulate a minimalistic model for the interacting electrons on honeycomb lattice described by the Hubbard Hamiltonian with hoppings extending up to the $5th$ nn. Namely, we take into account only the on-site $U$ interactions, as the inter-site interactions are---as mentioned---at least of two orders of magnitude weaker. Calculations have been carried out for the lattice consisting of $N=12\times12$ unit cells, that is, for $288$ sites within imposed periodic boundary conditions. 
The simulations have been carried out using the mVMC software~\cite{Misawa} with the help of self-elaborated, re-usable and generic Hamiltonian input generator~\cite{andrzej_biborski_zenodo}. The trial state is in Pfaffian form, i.e. product of antiparallel spin pairing terms, supplied with the on-site Gutzwiller and inter-site long-range Jastrow projectors, which are responsible for capturing the electronic correlation effects~\cite{Biborski0}. Variational parameters corresponding to pairings, Gutzwiller and Jastrow projectors are chosen with periodicity defined by a supercell consisting of $2\times2$ unit cells. 
%That is, for the $12\times12$ lattice, it results in $8$ Gutzwiller factors, $1160$ Jastrow coefficients, and $2304$ unique pairings when periodic boundary conditions are imposed. 
According to the spin polarization of the system, we impose the constraint $\sum_{i}^{2N}\langle \hat{S}^z_i \rangle = 0$, that is, the total $z$ component of the spin is zero.

In the following, we present the calculations obtained for $\lambda_{TF}=2.5$ nm, emphasizing that it can be regarded as overestimated for the strongly confined regime. However, as will become clear, it allows us to reveal typical features of the strongly correlated system when $r_0>19$ nm.

First, we analyze the average double occupancy defined as
\begin{align}
    \langle \hat{d}  \rangle \equiv \frac{1}{2N}\sum_{i=1}^{2N}\langle \hat{n}_{i\uparrow}\hat{n}_{i\downarrow}\rangle,
\end{align}
where $\hat{n}_{i\sigma}$ is the particle number operator associated with the site labeled by $i$ and $\sigma=\{\uparrow,\downarrow\}$ is the spin component $z$ of the electron that resides at the site $i$. 
\begin{figure}
\includegraphics[width=0.5\textwidth]{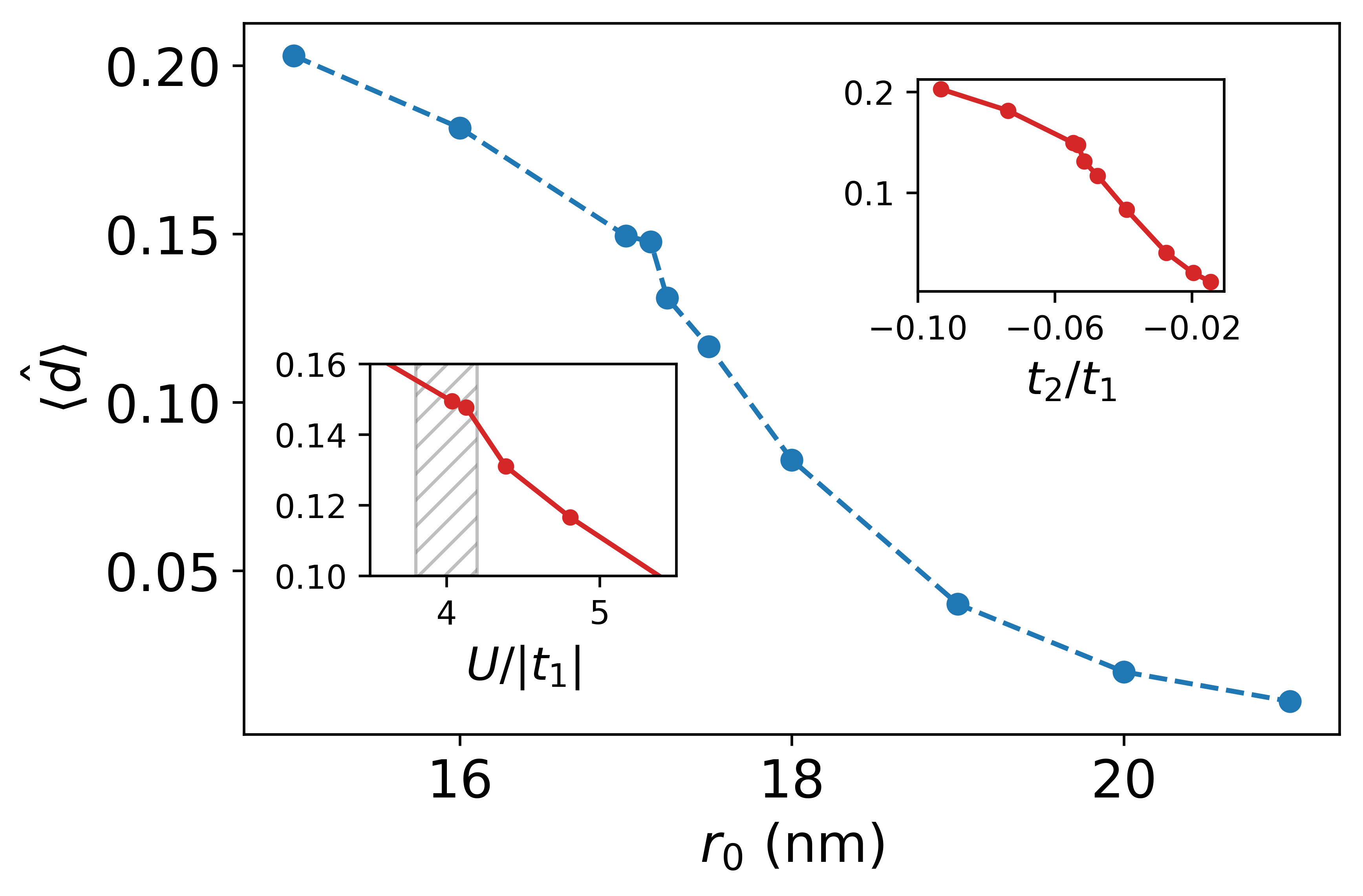}
\caption{The double occupancy $\langle\hat{d}\rangle$ as a function of the AD radius $r_0$ (main figure) and with respect to $U/|t_1|$ and $t2/t1$ in the inset figures. The estimated statistical errors are smaller than the symbol size. The dashed area in lower-left inset corresponds to $U_c\approx3.8\div4.2$, i.e., estimated critical value for which the AFMI state emerges.} 
\label{fig:doubleocc}
\end{figure}
In Fig.~\ref{fig:doubleocc} we show $\langle \hat{d} \rangle$ as a function of $r_0$. Its value decreases with an increasing AD radius. This behavior is driven by the mutual enhancement of $U$, as well as the narrowing of the occupied band. We performed auxiliary calculations for $r_0=17.15, 17.25$ and $17.5$ nm to identify  an abrupt drop in $\langle\hat{d}\rangle$ in vicinity of $r_0\approx17$ nm, which is the characteristic indicator of transition to the insulating state driven by electronic correlations. 

The emergence of AFMI state is also supported by the analysis of the spin-spin correlation functions. Namely, we investigate the AF spin order parameter provided by Sorella et al.~\cite{Sorella2012}, but defined for  the $z$-component of $\hat{S}_i$, i.e.,
\begin{align}
m_{s^z}^2\equiv\frac{1}{4N^2}\Bigg\langle\Big[\sum_{ij}\big(\hat{S}^z_{i,\alpha}-\hat{S}^z_{j,\beta}\big)\Big]^2 \Bigg\rangle.
\label{eq:msz}
\end{align}
\begin{figure}
\includegraphics[width=0.5\textwidth]{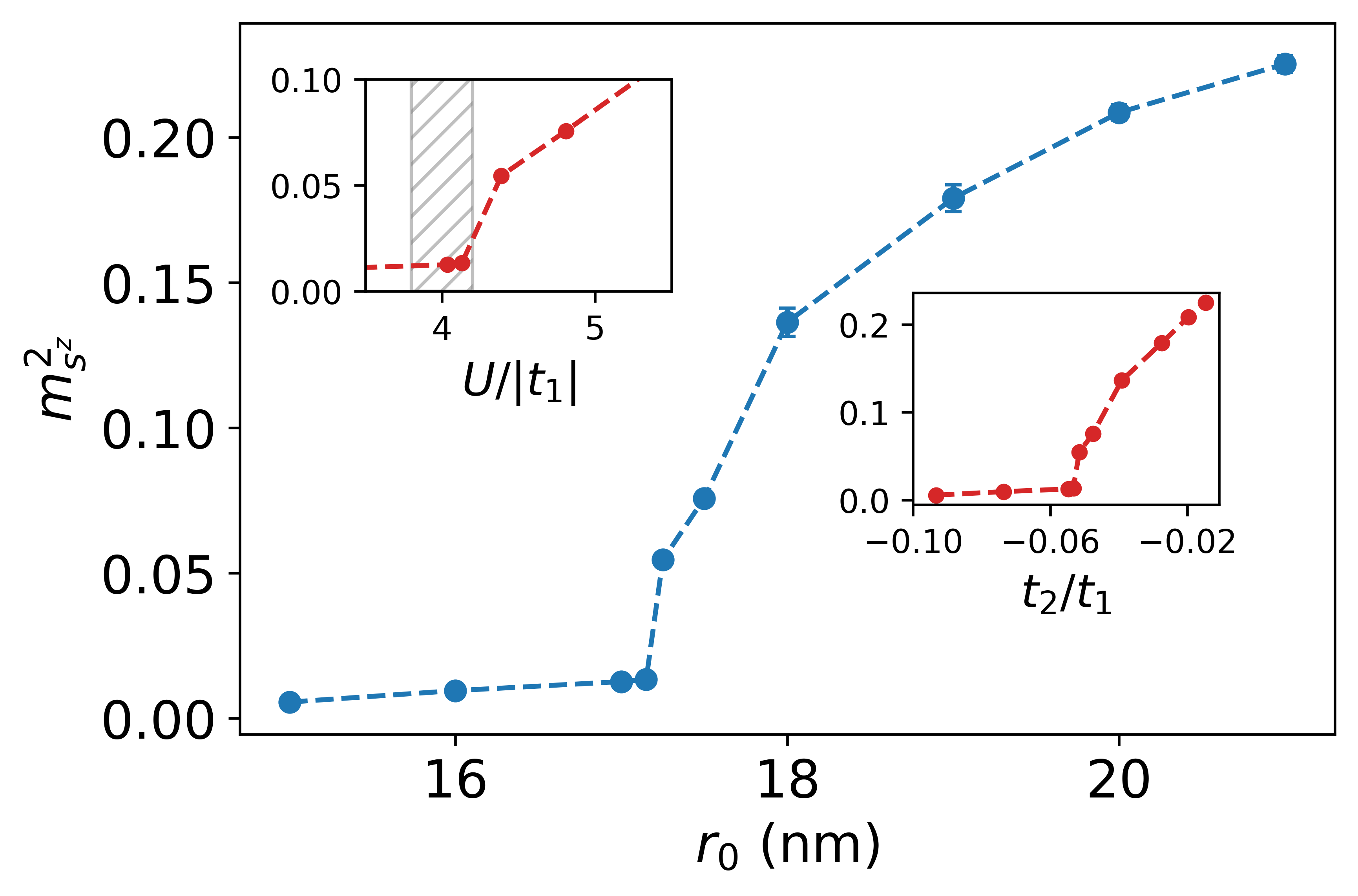}
\caption{The spin order parameter $m_{s^z}^2$ defined in Eq.~\ref{eq:msz} as a function of the AD radius (main figure). The insets contain $m_{s^z}^2$ dependencies on $U/|t_1|$ and $t_2/t_1$. The dashed area in upper-left inset corresponds to $U_c\approx3.8\div4.2$, i.e., estimated critical value for which the AFMI state appears with increasing value of $U$. Estimated statistical errors are smaller than the size of the symbols.}
\label{fig:spinaf}
\end{figure}

In Fig.~\ref{fig:spinaf} we present $m_{s^z}^2$ analogously to the double occupancies shown in Fig.~\ref{fig:doubleocc}. It is clearly visible that the AF order for $r_0\leq17$ nm is marginal since $m_{s^z}^2\propto10^{-3}$. However, for $r_0>17$ nm, we observe its radical increase, namely, for $r_0=18$ nm, which is the next case considered, it is $\sim 0.14$. 

Recapitulating, when $r_0$ is translated into the ratio $U/|t_1|$ (see the insets in Figs.~\ref{fig:doubleocc},~\ref{fig:spinaf}), simultaneous, anomal behaviour of both $\langle\hat{d}\rangle$ and $m_{s^z}^2$ can be observed for $r_0=17.15\div17.25$ nm. These values of $r_0$  correspond to $U/|t_1|=4.13\div4.38$, the range which nearly overlap with $3.8\div4.2$, i.e., critical values for which transition to AFMI takes place, as it has been estimated in the framework of Quantum Monte-Carlo approaches~\cite{Sorella2012,Meng2010,Raczkowski}.

\section{Summary and conclusions}
In this paper, we studied the anti-dot honeycomb lattice model in view of the importance of electron-electron interactions. Within the assumed form of the quantum dot potential and the multilevel computational scheme, we provided evidence of the possibility of the formation of an antiferromagnetic Mott insulator. Our modeling scheme reveals that the radius of the antidot is crucial in this context. In addition, the inclusion of metallic screening among electrons seems to be essential for the strength of the electron-electron interactions, specifically when the radius of the antidot is small. Note that we did not have to take into account the screening of the metallic gates that induce the periodic potential~\cite{Vandersypen,Biborski0,Byrnes} as the honeycomb structure here results from the triangular etching pattern of the upper layers. Also, we do not take into account any kind of scattering centers, such as ionic impurities or any other kind of disorder in the system. Therefore, the whole screening considered here has its origin in the \emph{metallic} nature of 2DEGs~\cite{Kollar2002}. In this spirit our supporting VMC analysis reveals possible mapping of interacting electrons in AD honeycomb lattice to the Hubbard model.
Our study provides a straightforward way to exploit the devices that are currently fabricated for the realization of strongly correlated artificial Dirac systems. The experimental confirmation of our predictions would open up the opportunity for a better understanding of the electronic properties of graphene-like systems in a controllable manner. Moreover, we indicate that in addition to the correlation effects studied here, the AD honeycomb lattice formed in 2DEG may also serve as a simulation platform for strain-induced opening of the gap in honeycomb systems~\cite{Cocco2010,Rut2023}, since strain can be emulated by the distorted QD assembly. Thus, the proposed modeling scheme may also be considered useful from this point of view.

\section{Acknowledgement}
This work was supported by National Science Centre (NCN) agreement number UMO-2020/38/E/ST3/00418.

\bibliography{biborski_et_al}% Produces the bibliography via BibTeX.

\begin{thebibliography}{35}%
\makeatletter
\providecommand \@ifxundefined [1]{%
 \@ifx{#1\undefined}
}%
\providecommand \@ifnum [1]{%
 \ifnum #1\expandafter \@firstoftwo
 \else \expandafter \@secondoftwo
 \fi
}%
\providecommand \@ifx [1]{%
 \ifx #1\expandafter \@firstoftwo
 \else \expandafter \@secondoftwo
 \fi
}%
\providecommand \natexlab [1]{#1}%
\providecommand \enquote  [1]{``#1''}%
\providecommand \bibnamefont  [1]{#1}%
\providecommand \bibfnamefont [1]{#1}%
\providecommand \citenamefont [1]{#1}%
\providecommand \href@noop [0]{\@secondoftwo}%
\providecommand \href [0]{\begingroup \@sanitize@url \@href}%
\providecommand \@href[1]{\@@startlink{#1}\@@href}%
\providecommand \@@href[1]{\endgroup#1\@@endlink}%
\providecommand \@sanitize@url [0]{\catcode `\\12\catcode `\$12\catcode
  `\&12\catcode `\#12\catcode `\^12\catcode `\_12\catcode `\%12\relax}%
\providecommand \@@startlink[1]{}%
\providecommand \@@endlink[0]{}%
\providecommand \url  [0]{\begingroup\@sanitize@url \@url }%
\providecommand \@url [1]{\endgroup\@href {#1}{\urlprefix }}%
\providecommand \urlprefix  [0]{URL }%
\providecommand \Eprint [0]{\href }%
\providecommand \doibase [0]{http://dx.doi.org/}%
\providecommand \selectlanguage [0]{\@gobble}%
\providecommand \bibinfo  [0]{\@secondoftwo}%
\providecommand \bibfield  [0]{\@secondoftwo}%
\providecommand \translation [1]{[#1]}%
\providecommand \BibitemOpen [0]{}%
\providecommand \bibitemStop [0]{}%
\providecommand \bibitemNoStop [0]{.\EOS\space}%
\providecommand \EOS [0]{\spacefactor3000\relax}%
\providecommand \BibitemShut  [1]{\csname bibitem#1\endcsname}%
\let\auto@bib@innerbib\@empty
%</preamble>
\bibitem [{\citenamefont {Kotov}\ \emph {et~al.}(2012)\citenamefont {Kotov},
  \citenamefont {Uchoa}, \citenamefont {Pereira}, \citenamefont {Guinea},\ and\
  \citenamefont {Castro~Neto}}]{Kotov}%
  \BibitemOpen
  \bibfield  {author} {\bibinfo {author} {\bibfnamefont {V.~N.}\ \bibnamefont
  {Kotov}}, \bibinfo {author} {\bibfnamefont {B.}~\bibnamefont {Uchoa}},
  \bibinfo {author} {\bibfnamefont {V.~M.}\ \bibnamefont {Pereira}}, \bibinfo
  {author} {\bibfnamefont {F.}~\bibnamefont {Guinea}}, \ and\ \bibinfo {author}
  {\bibfnamefont {A.~H.}\ \bibnamefont {Castro~Neto}},\ }\href {\doibase
  10.1103/RevModPhys.84.1067} {\bibfield  {journal} {\bibinfo  {journal} {Rev.
  Mod. Phys.}\ }\textbf {\bibinfo {volume} {84}},\ \bibinfo {pages} {1067}
  (\bibinfo {year} {2012})}\BibitemShut {NoStop}%
\bibitem [{\citenamefont {Tang}\ \emph {et~al.}(2018)\citenamefont {Tang},
  \citenamefont {Leaw}, \citenamefont {Rodrigues}, \citenamefont {Herbut},
  \citenamefont {Sengupta}, \citenamefont {Assaad},\ and\ \citenamefont
  {Adam}}]{hokin}%
  \BibitemOpen
  \bibfield  {author} {\bibinfo {author} {\bibfnamefont {H.-K.}\ \bibnamefont
  {Tang}}, \bibinfo {author} {\bibfnamefont {J.~N.}\ \bibnamefont {Leaw}},
  \bibinfo {author} {\bibfnamefont {J.~N.~B.}\ \bibnamefont {Rodrigues}},
  \bibinfo {author} {\bibfnamefont {I.~F.}\ \bibnamefont {Herbut}}, \bibinfo
  {author} {\bibfnamefont {P.}~\bibnamefont {Sengupta}}, \bibinfo {author}
  {\bibfnamefont {F.~F.}\ \bibnamefont {Assaad}}, \ and\ \bibinfo {author}
  {\bibfnamefont {S.}~\bibnamefont {Adam}},\ }\href {\doibase
  10.1126/science.aao2934} {\bibfield  {journal} {\bibinfo  {journal}
  {Science}\ }\textbf {\bibinfo {volume} {361}},\ \bibinfo {pages} {570}
  (\bibinfo {year} {2018})},\ \Eprint
  {http://arxiv.org/abs/https://www.science.org/doi/pdf/10.1126/science.aao2934}
  {https://www.science.org/doi/pdf/10.1126/science.aao2934} \BibitemShut
  {NoStop}%
\bibitem [{\citenamefont {Das~Sarma}\ \emph {et~al.}(2007)\citenamefont
  {Das~Sarma}, \citenamefont {Hwang},\ and\ \citenamefont
  {Tse}}]{DasSarma2007}%
  \BibitemOpen
  \bibfield  {author} {\bibinfo {author} {\bibfnamefont {S.}~\bibnamefont
  {Das~Sarma}}, \bibinfo {author} {\bibfnamefont {E.~H.}\ \bibnamefont
  {Hwang}}, \ and\ \bibinfo {author} {\bibfnamefont {W.-K.}\ \bibnamefont
  {Tse}},\ }\href {\doibase 10.1103/PhysRevB.75.121406} {\bibfield  {journal}
  {\bibinfo  {journal} {Phys. Rev. B}\ }\textbf {\bibinfo {volume} {75}},\
  \bibinfo {pages} {121406} (\bibinfo {year} {2007})}\BibitemShut {NoStop}%
\bibitem [{\citenamefont {Wu}\ and\ \citenamefont {Tremblay}(2014)}]{WuWei}%
  \BibitemOpen
  \bibfield  {author} {\bibinfo {author} {\bibfnamefont {W.}~\bibnamefont
  {Wu}}\ and\ \bibinfo {author} {\bibfnamefont {A.-M.~S.}\ \bibnamefont
  {Tremblay}},\ }\href {\doibase 10.1103/PhysRevB.89.205128} {\bibfield
  {journal} {\bibinfo  {journal} {Phys. Rev. B}\ }\textbf {\bibinfo {volume}
  {89}},\ \bibinfo {pages} {205128} (\bibinfo {year} {2014})}\BibitemShut
  {NoStop}%
\bibitem [{\citenamefont {Assaad}\ and\ \citenamefont {Herbut}(2013)}]{Assad}%
  \BibitemOpen
  \bibfield  {author} {\bibinfo {author} {\bibfnamefont {F.~F.}\ \bibnamefont
  {Assaad}}\ and\ \bibinfo {author} {\bibfnamefont {I.~F.}\ \bibnamefont
  {Herbut}},\ }\href {\doibase 10.1103/PhysRevX.3.031010} {\bibfield  {journal}
  {\bibinfo  {journal} {Phys. Rev. X}\ }\textbf {\bibinfo {volume} {3}},\
  \bibinfo {pages} {031010} (\bibinfo {year} {2013})}\BibitemShut {NoStop}%
\bibitem [{\citenamefont {Sorella}\ \emph {et~al.}(2012)\citenamefont
  {Sorella}, \citenamefont {Otsuka},\ and\ \citenamefont
  {Yunoki}}]{Sorella2012}%
  \BibitemOpen
  \bibfield  {author} {\bibinfo {author} {\bibfnamefont {S.}~\bibnamefont
  {Sorella}}, \bibinfo {author} {\bibfnamefont {Y.}~\bibnamefont {Otsuka}}, \
  and\ \bibinfo {author} {\bibfnamefont {S.}~\bibnamefont {Yunoki}},\ }\href
  {\doibase 10.1038/srep00992} {\bibfield  {journal} {\bibinfo  {journal}
  {Scientific Reports}\ }\textbf {\bibinfo {volume} {2}},\ \bibinfo {pages}
  {992} (\bibinfo {year} {2012})}\BibitemShut {NoStop}%
\bibitem [{\citenamefont {Meng}\ \emph {et~al.}(2010)\citenamefont {Meng},
  \citenamefont {Lang}, \citenamefont {Wessel}, \citenamefont {Assaad},\ and\
  \citenamefont {Muramatsu}}]{Meng2010}%
  \BibitemOpen
  \bibfield  {author} {\bibinfo {author} {\bibfnamefont {Z.~Y.}\ \bibnamefont
  {Meng}}, \bibinfo {author} {\bibfnamefont {T.~C.}\ \bibnamefont {Lang}},
  \bibinfo {author} {\bibfnamefont {S.}~\bibnamefont {Wessel}}, \bibinfo
  {author} {\bibfnamefont {F.~F.}\ \bibnamefont {Assaad}}, \ and\ \bibinfo
  {author} {\bibfnamefont {A.}~\bibnamefont {Muramatsu}},\ }\href {\doibase
  10.1038/nature08942} {\bibfield  {journal} {\bibinfo  {journal} {Nature}\
  }\textbf {\bibinfo {volume} {464}},\ \bibinfo {pages} {847} (\bibinfo {year}
  {2010})}\BibitemShut {NoStop}%
\bibitem [{\citenamefont {Raczkowski}\ \emph {et~al.}(2020)\citenamefont
  {Raczkowski}, \citenamefont {Peters}, \citenamefont {Ph\`ung}, \citenamefont
  {Takemori}, \citenamefont {Assaad}, \citenamefont {Honecker},\ and\
  \citenamefont {Vahedi}}]{Raczkowski}%
  \BibitemOpen
  \bibfield  {author} {\bibinfo {author} {\bibfnamefont {M.}~\bibnamefont
  {Raczkowski}}, \bibinfo {author} {\bibfnamefont {R.}~\bibnamefont {Peters}},
  \bibinfo {author} {\bibfnamefont {T.~T.}\ \bibnamefont {Ph\`ung}}, \bibinfo
  {author} {\bibfnamefont {N.}~\bibnamefont {Takemori}}, \bibinfo {author}
  {\bibfnamefont {F.~F.}\ \bibnamefont {Assaad}}, \bibinfo {author}
  {\bibfnamefont {A.}~\bibnamefont {Honecker}}, \ and\ \bibinfo {author}
  {\bibfnamefont {J.}~\bibnamefont {Vahedi}},\ }\href {\doibase
  10.1103/PhysRevB.101.125103} {\bibfield  {journal} {\bibinfo  {journal}
  {Phys. Rev. B}\ }\textbf {\bibinfo {volume} {101}},\ \bibinfo {pages}
  {125103} (\bibinfo {year} {2020})}\BibitemShut {NoStop}%
\bibitem [{\citenamefont {Otsuka}\ \emph {et~al.}(2016)\citenamefont {Otsuka},
  \citenamefont {Yunoki},\ and\ \citenamefont {Sorella}}]{Otsuka}%
  \BibitemOpen
  \bibfield  {author} {\bibinfo {author} {\bibfnamefont {Y.}~\bibnamefont
  {Otsuka}}, \bibinfo {author} {\bibfnamefont {S.}~\bibnamefont {Yunoki}}, \
  and\ \bibinfo {author} {\bibfnamefont {S.}~\bibnamefont {Sorella}},\ }\href
  {\doibase 10.1103/PhysRevX.6.011029} {\bibfield  {journal} {\bibinfo
  {journal} {Phys. Rev. X}\ }\textbf {\bibinfo {volume} {6}},\ \bibinfo {pages}
  {011029} (\bibinfo {year} {2016})}\BibitemShut {NoStop}%
\bibitem [{\citenamefont {Liebsch}\ and\ \citenamefont {Wu}(2013)}]{Liebsch}%
  \BibitemOpen
  \bibfield  {author} {\bibinfo {author} {\bibfnamefont {A.}~\bibnamefont
  {Liebsch}}\ and\ \bibinfo {author} {\bibfnamefont {W.}~\bibnamefont {Wu}},\
  }\href {\doibase 10.1103/PhysRevB.87.205127} {\bibfield  {journal} {\bibinfo
  {journal} {Phys. Rev. B}\ }\textbf {\bibinfo {volume} {87}},\ \bibinfo
  {pages} {205127} (\bibinfo {year} {2013})}\BibitemShut {NoStop}%
\bibitem [{\citenamefont {Arya}\ \emph {et~al.}(2015)\citenamefont {Arya},
  \citenamefont {Sriluckshmy}, \citenamefont {Hassan},\ and\ \citenamefont
  {Tremblay}}]{Arya}%
  \BibitemOpen
  \bibfield  {author} {\bibinfo {author} {\bibfnamefont {S.}~\bibnamefont
  {Arya}}, \bibinfo {author} {\bibfnamefont {P.~V.}\ \bibnamefont
  {Sriluckshmy}}, \bibinfo {author} {\bibfnamefont {S.~R.}\ \bibnamefont
  {Hassan}}, \ and\ \bibinfo {author} {\bibfnamefont {A.-M.~S.}\ \bibnamefont
  {Tremblay}},\ }\href {\doibase 10.1103/PhysRevB.92.045111} {\bibfield
  {journal} {\bibinfo  {journal} {Phys. Rev. B}\ }\textbf {\bibinfo {volume}
  {92}},\ \bibinfo {pages} {045111} (\bibinfo {year} {2015})}\BibitemShut
  {NoStop}%
\bibitem [{\citenamefont {Saleem}\ \emph {et~al.}(2022)\citenamefont {Saleem},
  \citenamefont {Dusko}, \citenamefont {Cygorek}, \citenamefont {Korkusinski},\
  and\ \citenamefont {Hawrylak}}]{Saleem}%
  \BibitemOpen
  \bibfield  {author} {\bibinfo {author} {\bibfnamefont {Y.}~\bibnamefont
  {Saleem}}, \bibinfo {author} {\bibfnamefont {A.}~\bibnamefont {Dusko}},
  \bibinfo {author} {\bibfnamefont {M.}~\bibnamefont {Cygorek}}, \bibinfo
  {author} {\bibfnamefont {M.}~\bibnamefont {Korkusinski}}, \ and\ \bibinfo
  {author} {\bibfnamefont {P.}~\bibnamefont {Hawrylak}},\ }\href {\doibase
  10.1103/PhysRevB.105.205105} {\bibfield  {journal} {\bibinfo  {journal}
  {Phys. Rev. B}\ }\textbf {\bibinfo {volume} {105}},\ \bibinfo {pages}
  {205105} (\bibinfo {year} {2022})}\BibitemShut {NoStop}%
\bibitem [{\citenamefont {Du}\ \emph {et~al.}(2021)\citenamefont {Du},
  \citenamefont {Liu}, \citenamefont {Wind}, \citenamefont {Pellegrini},
  \citenamefont {West}, \citenamefont {Fallahi}, \citenamefont {Pfeiffer},
  \citenamefont {Manfra},\ and\ \citenamefont {Pinczuk}}]{DuLiu}%
  \BibitemOpen
  \bibfield  {author} {\bibinfo {author} {\bibfnamefont {L.}~\bibnamefont
  {Du}}, \bibinfo {author} {\bibfnamefont {Z.}~\bibnamefont {Liu}}, \bibinfo
  {author} {\bibfnamefont {S.~J.}\ \bibnamefont {Wind}}, \bibinfo {author}
  {\bibfnamefont {V.}~\bibnamefont {Pellegrini}}, \bibinfo {author}
  {\bibfnamefont {K.~W.}\ \bibnamefont {West}}, \bibinfo {author}
  {\bibfnamefont {S.}~\bibnamefont {Fallahi}}, \bibinfo {author} {\bibfnamefont
  {L.~N.}\ \bibnamefont {Pfeiffer}}, \bibinfo {author} {\bibfnamefont {M.~J.}\
  \bibnamefont {Manfra}}, \ and\ \bibinfo {author} {\bibfnamefont
  {A.}~\bibnamefont {Pinczuk}},\ }\href {\doibase
  10.1103/PhysRevLett.126.106402} {\bibfield  {journal} {\bibinfo  {journal}
  {Phys. Rev. Lett.}\ }\textbf {\bibinfo {volume} {126}},\ \bibinfo {pages}
  {106402} (\bibinfo {year} {2021})}\BibitemShut {NoStop}%
\bibitem [{\citenamefont {Kylänpää}\ \emph {et~al.}(2016)\citenamefont
  {Kylänpää}, \citenamefont {Berardi}, \citenamefont {Räsänen},
  \citenamefont {García-González}, \citenamefont {Rozzi},\ and\ \citenamefont
  {Rubio}}]{Kylanpaa}%
  \BibitemOpen
  \bibfield  {author} {\bibinfo {author} {\bibfnamefont {I.}~\bibnamefont
  {Kylänpää}}, \bibinfo {author} {\bibfnamefont {F.}~\bibnamefont
  {Berardi}}, \bibinfo {author} {\bibfnamefont {E.}~\bibnamefont {Räsänen}},
  \bibinfo {author} {\bibfnamefont {P.}~\bibnamefont {García-González}},
  \bibinfo {author} {\bibfnamefont {C.~A.}\ \bibnamefont {Rozzi}}, \ and\
  \bibinfo {author} {\bibfnamefont {A.}~\bibnamefont {Rubio}},\ }\href
  {\doibase 10.1088/1367-2630/18/8/083014} {\bibfield  {journal} {\bibinfo
  {journal} {New J. Phys.}\ }\textbf {\bibinfo {volume} {18}} (\bibinfo {year}
  {2016}),\ 10.1088/1367-2630/18/8/083014}\BibitemShut {NoStop}%
\bibitem [{\citenamefont {Li}\ \emph {et~al.}(2020)\citenamefont {Li},
  \citenamefont {Ingham},\ and\ \citenamefont {Scammell}}]{TommyLi}%
  \BibitemOpen
  \bibfield  {author} {\bibinfo {author} {\bibfnamefont {T.}~\bibnamefont
  {Li}}, \bibinfo {author} {\bibfnamefont {J.}~\bibnamefont {Ingham}}, \ and\
  \bibinfo {author} {\bibfnamefont {H.~D.}\ \bibnamefont {Scammell}},\ }\href
  {\doibase 10.1103/PhysRevResearch.2.043155} {\bibfield  {journal} {\bibinfo
  {journal} {Phys. Rev. Res.}\ }\textbf {\bibinfo {volume} {2}},\ \bibinfo
  {pages} {043155} (\bibinfo {year} {2020})}\BibitemShut {NoStop}%
\bibitem [{\citenamefont {R\"as\"anen}\ \emph {et~al.}(2012)\citenamefont
  {R\"as\"anen}, \citenamefont {Rozzi}, \citenamefont {Pittalis},\ and\
  \citenamefont {Vignale}}]{Rasanen}%
  \BibitemOpen
  \bibfield  {author} {\bibinfo {author} {\bibfnamefont {E.}~\bibnamefont
  {R\"as\"anen}}, \bibinfo {author} {\bibfnamefont {C.~A.}\ \bibnamefont
  {Rozzi}}, \bibinfo {author} {\bibfnamefont {S.}~\bibnamefont {Pittalis}}, \
  and\ \bibinfo {author} {\bibfnamefont {G.}~\bibnamefont {Vignale}},\ }\href
  {\doibase 10.1103/PhysRevLett.108.246803} {\bibfield  {journal} {\bibinfo
  {journal} {Phys. Rev. Lett.}\ }\textbf {\bibinfo {volume} {108}},\ \bibinfo
  {pages} {246803} (\bibinfo {year} {2012})}\BibitemShut {NoStop}%
\bibitem [{\citenamefont {Krix}\ \emph {et~al.}(2022)\citenamefont {Krix},
  \citenamefont {Scammell},\ and\ \citenamefont {Sushkov}}]{Krix}%
  \BibitemOpen
  \bibfield  {author} {\bibinfo {author} {\bibfnamefont {Z.~E.}\ \bibnamefont
  {Krix}}, \bibinfo {author} {\bibfnamefont {H.~D.}\ \bibnamefont {Scammell}},
  \ and\ \bibinfo {author} {\bibfnamefont {O.~P.}\ \bibnamefont {Sushkov}},\
  }\href {\doibase 10.1103/PhysRevB.105.075120} {\bibfield  {journal} {\bibinfo
   {journal} {Phys. Rev. B}\ }\textbf {\bibinfo {volume} {105}},\ \bibinfo
  {pages} {075120} (\bibinfo {year} {2022})}\BibitemShut {NoStop}%
\bibitem [{\citenamefont {Li}\ \emph {et~al.}(2017)\citenamefont {Li},
  \citenamefont {Zhou}, \citenamefont {Zhang}, \citenamefont {Zhang},\ and\
  \citenamefont {Chang}}]{Yun-Mei}%
  \BibitemOpen
  \bibfield  {author} {\bibinfo {author} {\bibfnamefont {Y.-M.}\ \bibnamefont
  {Li}}, \bibinfo {author} {\bibfnamefont {X.}~\bibnamefont {Zhou}}, \bibinfo
  {author} {\bibfnamefont {Y.-Y.}\ \bibnamefont {Zhang}}, \bibinfo {author}
  {\bibfnamefont {D.}~\bibnamefont {Zhang}}, \ and\ \bibinfo {author}
  {\bibfnamefont {K.}~\bibnamefont {Chang}},\ }\href {\doibase
  10.1103/PhysRevB.96.035406} {\bibfield  {journal} {\bibinfo  {journal} {Phys.
  Rev. B}\ }\textbf {\bibinfo {volume} {96}},\ \bibinfo {pages} {035406}
  (\bibinfo {year} {2017})}\BibitemShut {NoStop}%
\bibitem [{\citenamefont {Tkachenko}\ \emph {et~al.}(2015)\citenamefont
  {Tkachenko}, \citenamefont {Tkachenko}, \citenamefont {Terekhov},\ and\
  \citenamefont {Sushkov}}]{Tkachenko_2015}%
  \BibitemOpen
  \bibfield  {author} {\bibinfo {author} {\bibfnamefont {O.~A.}\ \bibnamefont
  {Tkachenko}}, \bibinfo {author} {\bibfnamefont {V.~A.}\ \bibnamefont
  {Tkachenko}}, \bibinfo {author} {\bibfnamefont {I.~S.}\ \bibnamefont
  {Terekhov}}, \ and\ \bibinfo {author} {\bibfnamefont {O.~P.}\ \bibnamefont
  {Sushkov}},\ }\href {\doibase 10.1088/2053-1583/2/1/014010} {\bibfield
  {journal} {\bibinfo  {journal} {2D Materials}\ }\textbf {\bibinfo {volume}
  {2}},\ \bibinfo {pages} {014010} (\bibinfo {year} {2015})}\BibitemShut
  {NoStop}%
\bibitem [{\citenamefont {Bednarek}\ \emph {et~al.}(2003)\citenamefont
  {Bednarek}, \citenamefont {Szafran}, \citenamefont {Lis},\ and\ \citenamefont
  {Adamowski}}]{Bednarek2}%
  \BibitemOpen
  \bibfield  {author} {\bibinfo {author} {\bibfnamefont {S.}~\bibnamefont
  {Bednarek}}, \bibinfo {author} {\bibfnamefont {B.}~\bibnamefont {Szafran}},
  \bibinfo {author} {\bibfnamefont {K.}~\bibnamefont {Lis}}, \ and\ \bibinfo
  {author} {\bibfnamefont {J.}~\bibnamefont {Adamowski}},\ }\href {\doibase
  10.1103/PhysRevB.68.155333} {\bibfield  {journal} {\bibinfo  {journal} {Phys.
  Rev. B}\ }\textbf {\bibinfo {volume} {68}},\ \bibinfo {pages} {155333}
  (\bibinfo {year} {2003})}\BibitemShut {NoStop}%
\bibitem [{\citenamefont {Ciurla}\ \emph {et~al.}(2002)\citenamefont {Ciurla},
  \citenamefont {Adamowski}, \citenamefont {Szafran},\ and\ \citenamefont
  {Bednarek}}]{Ciurla}%
  \BibitemOpen
  \bibfield  {author} {\bibinfo {author} {\bibfnamefont {M.}~\bibnamefont
  {Ciurla}}, \bibinfo {author} {\bibfnamefont {J.}~\bibnamefont {Adamowski}},
  \bibinfo {author} {\bibfnamefont {B.}~\bibnamefont {Szafran}}, \ and\
  \bibinfo {author} {\bibfnamefont {S.}~\bibnamefont {Bednarek}},\ }\href
  {\doibase https://doi.org/10.1016/S1386-9477(02)00572-6} {\bibfield
  {journal} {\bibinfo  {journal} {Physica E Low Dimens. Syst. Nanostruct.}\
  }\textbf {\bibinfo {volume} {15}},\ \bibinfo {pages} {261} (\bibinfo {year}
  {2002})}\BibitemShut {NoStop}%
\bibitem [{Note1()}]{Note1}%
  \BibitemOpen
  \bibinfo {note} {Note that we excluded the bottom layer of $n$-doped AlGaAs
  as detailed information regarding the character of doping has not been
  provided ~\cite {DuLiu}}\BibitemShut {NoStop}%
\bibitem [{\citenamefont {Marzari}\ \emph {et~al.}(2012)\citenamefont
  {Marzari}, \citenamefont {Mostofi}, \citenamefont {Yates}, \citenamefont
  {Souza},\ and\ \citenamefont {Vanderbilt}}]{Marzari}%
  \BibitemOpen
  \bibfield  {author} {\bibinfo {author} {\bibfnamefont {N.}~\bibnamefont
  {Marzari}}, \bibinfo {author} {\bibfnamefont {A.~A.}\ \bibnamefont
  {Mostofi}}, \bibinfo {author} {\bibfnamefont {J.~R.}\ \bibnamefont {Yates}},
  \bibinfo {author} {\bibfnamefont {I.}~\bibnamefont {Souza}}, \ and\ \bibinfo
  {author} {\bibfnamefont {D.}~\bibnamefont {Vanderbilt}},\ }\href {\doibase
  10.1103/RevModPhys.84.1419} {\bibfield  {journal} {\bibinfo  {journal} {Rev.
  Mod. Phys.}\ }\textbf {\bibinfo {volume} {84}},\ \bibinfo {pages} {1419}
  (\bibinfo {year} {2012})}\BibitemShut {NoStop}%
\bibitem [{\citenamefont {Groth}\ \emph {et~al.}(2014)\citenamefont {Groth},
  \citenamefont {Wimmer}, \citenamefont {Akhmerov},\ and\ \citenamefont
  {Waintal}}]{Kwant}%
  \BibitemOpen
  \bibfield  {author} {\bibinfo {author} {\bibfnamefont {C.~W.}\ \bibnamefont
  {Groth}}, \bibinfo {author} {\bibfnamefont {M.}~\bibnamefont {Wimmer}},
  \bibinfo {author} {\bibfnamefont {A.~R.}\ \bibnamefont {Akhmerov}}, \ and\
  \bibinfo {author} {\bibfnamefont {X.}~\bibnamefont {Waintal}},\ }\href
  {\doibase 10.1088/1367-2630/16/6/063065} {\bibfield  {journal} {\bibinfo
  {journal} {New J. Phys.}\ }\textbf {\bibinfo {volume} {16}},\ \bibinfo
  {pages} {063065} (\bibinfo {year} {2014})}\BibitemShut {NoStop}%
\bibitem [{\citenamefont {Biborski}\ \emph {et~al.}(2021)\citenamefont
  {Biborski}, \citenamefont {Nowak},\ and\ \citenamefont
  {Zegrodnik}}]{Biborski0}%
  \BibitemOpen
  \bibfield  {author} {\bibinfo {author} {\bibfnamefont {A.}~\bibnamefont
  {Biborski}}, \bibinfo {author} {\bibfnamefont {M.~P.}\ \bibnamefont {Nowak}},
  \ and\ \bibinfo {author} {\bibfnamefont {M.}~\bibnamefont {Zegrodnik}},\
  }\href {\doibase 10.1103/PhysRevB.104.245430} {\bibfield  {journal} {\bibinfo
   {journal} {Phys. Rev. B}\ }\textbf {\bibinfo {volume} {104}},\ \bibinfo
  {pages} {245430} (\bibinfo {year} {2021})}\BibitemShut {NoStop}%
\bibitem [{\citenamefont {Kn\"orzer}\ \emph {et~al.}(2022)\citenamefont
  {Kn\"orzer}, \citenamefont {van Diepen}, \citenamefont {Hsiao}, \citenamefont
  {Giedke}, \citenamefont {Mukhopadhyay}, \citenamefont {Reichl}, \citenamefont
  {Wegscheider}, \citenamefont {Cirac},\ and\ \citenamefont
  {Vandersypen}}]{Vandersypen}%
  \BibitemOpen
  \bibfield  {author} {\bibinfo {author} {\bibfnamefont {J.}~\bibnamefont
  {Kn\"orzer}}, \bibinfo {author} {\bibfnamefont {C.~J.}\ \bibnamefont {van
  Diepen}}, \bibinfo {author} {\bibfnamefont {T.-K.}\ \bibnamefont {Hsiao}},
  \bibinfo {author} {\bibfnamefont {G.}~\bibnamefont {Giedke}}, \bibinfo
  {author} {\bibfnamefont {U.}~\bibnamefont {Mukhopadhyay}}, \bibinfo {author}
  {\bibfnamefont {C.}~\bibnamefont {Reichl}}, \bibinfo {author} {\bibfnamefont
  {W.}~\bibnamefont {Wegscheider}}, \bibinfo {author} {\bibfnamefont {J.~I.}\
  \bibnamefont {Cirac}}, \ and\ \bibinfo {author} {\bibfnamefont {L.~M.~K.}\
  \bibnamefont {Vandersypen}},\ }\href {\doibase
  10.1103/PhysRevResearch.4.033043} {\bibfield  {journal} {\bibinfo  {journal}
  {Phys. Rev. Res.}\ }\textbf {\bibinfo {volume} {4}},\ \bibinfo {pages}
  {033043} (\bibinfo {year} {2022})}\BibitemShut {NoStop}%
\bibitem [{\citenamefont {Ortiz}\ \emph {et~al.}(1999)\citenamefont {Ortiz},
  \citenamefont {Harris},\ and\ \citenamefont {Ballone}}]{Ortiz}%
  \BibitemOpen
  \bibfield  {author} {\bibinfo {author} {\bibfnamefont {G.}~\bibnamefont
  {Ortiz}}, \bibinfo {author} {\bibfnamefont {M.}~\bibnamefont {Harris}}, \
  and\ \bibinfo {author} {\bibfnamefont {P.}~\bibnamefont {Ballone}},\ }\href
  {\doibase 10.1103/PhysRevLett.82.5317} {\bibfield  {journal} {\bibinfo
  {journal} {Phys. Rev. Lett.}\ }\textbf {\bibinfo {volume} {82}},\ \bibinfo
  {pages} {5317} (\bibinfo {year} {1999})}\BibitemShut {NoStop}%
\bibitem [{\citenamefont {Moreno}\ and\ \citenamefont
  {Méndez-Moreno}(2001)}]{Moreno}%
  \BibitemOpen
  \bibfield  {author} {\bibinfo {author} {\bibfnamefont {M.}~\bibnamefont
  {Moreno}}\ and\ \bibinfo {author} {\bibfnamefont {R.~M.}\ \bibnamefont
  {Méndez-Moreno}},\ }\href {\doibase https://doi.org/10.1002/qua.1049}
  {\bibfield  {journal} {\bibinfo  {journal} {International Journal of Quantum
  Chemistry}\ }\textbf {\bibinfo {volume} {82}},\ \bibinfo {pages} {269}
  (\bibinfo {year} {2001})},\ \Eprint
  {http://arxiv.org/abs/https://onlinelibrary.wiley.com/doi/pdf/10.1002/qua.1049}
  {https://onlinelibrary.wiley.com/doi/pdf/10.1002/qua.1049} \BibitemShut
  {NoStop}%
\bibitem [{\citenamefont {Hahn}(2005)}]{Cuba}%
  \BibitemOpen
  \bibfield  {author} {\bibinfo {author} {\bibfnamefont {T.}~\bibnamefont
  {Hahn}},\ }\href {\doibase https://doi.org/10.1016/j.cpc.2005.01.010}
  {\bibfield  {journal} {\bibinfo  {journal} {Comput. Phys. Commun.}\ }\textbf
  {\bibinfo {volume} {168}},\ \bibinfo {pages} {78} (\bibinfo {year}
  {2005})}\BibitemShut {NoStop}%
\bibitem [{\citenamefont {Misawa}\ \emph {et~al.}(2019)\citenamefont {Misawa},
  \citenamefont {Morita}, \citenamefont {Yoshimi}, \citenamefont {Kawamura},
  \citenamefont {Motoyama}, \citenamefont {Ido}, \citenamefont {Ohgoe},
  \citenamefont {Imada},\ and\ \citenamefont {Kato}}]{Misawa}%
  \BibitemOpen
  \bibfield  {author} {\bibinfo {author} {\bibfnamefont {T.}~\bibnamefont
  {Misawa}}, \bibinfo {author} {\bibfnamefont {S.}~\bibnamefont {Morita}},
  \bibinfo {author} {\bibfnamefont {K.}~\bibnamefont {Yoshimi}}, \bibinfo
  {author} {\bibfnamefont {M.}~\bibnamefont {Kawamura}}, \bibinfo {author}
  {\bibfnamefont {Y.}~\bibnamefont {Motoyama}}, \bibinfo {author}
  {\bibfnamefont {K.}~\bibnamefont {Ido}}, \bibinfo {author} {\bibfnamefont
  {T.}~\bibnamefont {Ohgoe}}, \bibinfo {author} {\bibfnamefont
  {M.}~\bibnamefont {Imada}}, \ and\ \bibinfo {author} {\bibfnamefont
  {T.}~\bibnamefont {Kato}},\ }\href {\doibase
  https://doi.org/10.1016/j.cpc.2018.08.014} {\bibfield  {journal} {\bibinfo
  {journal} {Comput. Phys. Commun.}\ }\textbf {\bibinfo {volume} {235}},\
  \bibinfo {pages} {447} (\bibinfo {year} {2019})}\BibitemShut {NoStop}%
\bibitem [{\citenamefont {Biborski}(2023)}]{andrzej_biborski_zenodo}%
  \BibitemOpen
  \bibfield  {author} {\bibinfo {author} {\bibfnamefont {A.}~\bibnamefont
  {Biborski}},\ }\href {\doibase 10.5281/zenodo.8172389} {\  (\bibinfo {year}
  {2023}),\ 10.5281/zenodo.8172389}\BibitemShut {NoStop}%
\bibitem [{\citenamefont {Byrnes}\ \emph {et~al.}(2008)\citenamefont {Byrnes},
  \citenamefont {Kim}, \citenamefont {Kusudo},\ and\ \citenamefont
  {Yamamoto}}]{Byrnes}%
  \BibitemOpen
  \bibfield  {author} {\bibinfo {author} {\bibfnamefont {T.}~\bibnamefont
  {Byrnes}}, \bibinfo {author} {\bibfnamefont {N.~Y.}\ \bibnamefont {Kim}},
  \bibinfo {author} {\bibfnamefont {K.}~\bibnamefont {Kusudo}}, \ and\ \bibinfo
  {author} {\bibfnamefont {Y.}~\bibnamefont {Yamamoto}},\ }\href {\doibase
  10.1103/PhysRevB.78.075320} {\bibfield  {journal} {\bibinfo  {journal} {Phys.
  Rev. B}\ }\textbf {\bibinfo {volume} {78}},\ \bibinfo {pages} {075320}
  (\bibinfo {year} {2008})}\BibitemShut {NoStop}%
\bibitem [{\citenamefont {Kollar}\ and\ \citenamefont
  {Sachdev}(2002)}]{Kollar2002}%
  \BibitemOpen
  \bibfield  {author} {\bibinfo {author} {\bibfnamefont {M.}~\bibnamefont
  {Kollar}}\ and\ \bibinfo {author} {\bibfnamefont {S.}~\bibnamefont
  {Sachdev}},\ }\href {\doibase 10.1103/PhysRevB.65.121304} {\bibfield
  {journal} {\bibinfo  {journal} {Phys. Rev. B}\ }\textbf {\bibinfo {volume}
  {65}},\ \bibinfo {pages} {121304} (\bibinfo {year} {2002})}\BibitemShut
  {NoStop}%
\bibitem [{\citenamefont {Cocco}\ \emph {et~al.}(2010)\citenamefont {Cocco},
  \citenamefont {Cadelano},\ and\ \citenamefont {Colombo}}]{Cocco2010}%
  \BibitemOpen
  \bibfield  {author} {\bibinfo {author} {\bibfnamefont {G.}~\bibnamefont
  {Cocco}}, \bibinfo {author} {\bibfnamefont {E.}~\bibnamefont {Cadelano}}, \
  and\ \bibinfo {author} {\bibfnamefont {L.}~\bibnamefont {Colombo}},\ }\href
  {\doibase 10.1103/PhysRevB.81.241412} {\bibfield  {journal} {\bibinfo
  {journal} {Phys. Rev. B}\ }\textbf {\bibinfo {volume} {81}},\ \bibinfo
  {pages} {241412} (\bibinfo {year} {2010})}\BibitemShut {NoStop}%
\bibitem [{\citenamefont {Rut}\ \emph {et~al.}(2023)\citenamefont {Rut},
  \citenamefont {Fidrysiak}, \citenamefont {Goc-Jagło},\ and\ \citenamefont
  {Rycerz}}]{Rut2023}%
  \BibitemOpen
  \bibfield  {author} {\bibinfo {author} {\bibfnamefont {G.}~\bibnamefont
  {Rut}}, \bibinfo {author} {\bibfnamefont {M.}~\bibnamefont {Fidrysiak}},
  \bibinfo {author} {\bibfnamefont {D.}~\bibnamefont {Goc-Jagło}}, \ and\
  \bibinfo {author} {\bibfnamefont {A.}~\bibnamefont {Rycerz}},\ }\href
  {\doibase 10.3390/ijms24021509} {\bibfield  {journal} {\bibinfo  {journal}
  {International Journal of Molecular Sciences}\ }\textbf {\bibinfo {volume}
  {24}} (\bibinfo {year} {2023}),\ 10.3390/ijms24021509}\BibitemShut {NoStop}%
\end{thebibliography}%

\end{document}